\documentclass[11pt]{article}%
\usepackage{graphicx}
\usepackage{amsfonts}
\usepackage{amsmath}
\usepackage{amssymb}
\usepackage{xcolor}
\usepackage{subcaption}%
\usepackage{natbib}[square,authoryear]
\setcitestyle{authoryear,open={(},close={)}} 

\usepackage{hyperref}
\providecommand{\U}[1]{\protect\rule{.1in}{.1in}}

\newtheorem{theorem}{Theorem}

\newtheorem{lemma}[theorem]{Lemma}

\newtheorem{remark}[theorem]{Remark}

\begin{document}


\title{\textbf{Robust adaptive Lasso in high-dimensional logistic regression.}}
\author{Basu, A.$^{1}$, Ghosh, A.$^{1}$; Jaenada, M.$^{2}$ and Pardo, L.$^{2}$\\$^{1}${\small Indian Statistical Institute, India}\\$^{2}${\small Complutense University of Madrid, Spain} }
\date{\today }
\maketitle


\begin{abstract}
Penalized logistic regression is extremely useful for binary classification with large number of covariates (higher than the sample size), having  several  real life applications, including genomic  disease classification. However, the existing methods based on the likelihood loss function are sensitive to data contamination and other noise and, hence, robust methods are needed for stable and more accurate inference.
In this paper, we propose a family of robust estimators for sparse logistic models utilizing the popular density power divergence based loss function and the general adaptively weighted LASSO penalties. We study the local robustness of the proposed estimators through its influence function and also derive its oracle properties and asymptotic distribution. With extensive empirical illustrations, we demonstrate the significantly improved performance of our proposed estimators over the existing ones with particular gain in robustness. Our proposal is finally applied to analyse  four different real datasets for cancer classification, obtaining robust and accurate models, that simultaneously performs  gene selection and patient classification.
\end{abstract}



\section{Introduction\label{sec1}}

Real life scientific experiments often include dichotomous response, which requires the use of binary classification procedures. Logistic regression can be used as a discriminative classification technique, having a direct probabilistic interpretation.  Let $Y_{1},...,Y_{n}$ be independent binary response variables following a Bernoulli model, i.e., 
\[
\Pr(Y_{i}=1)=\pi_{i}\text{ and }\Pr(Y_{i}=0)=1-\pi_{i},\text{ }i=1,...,n.
\]
Often, $\pi_{i} \in [0,1]$ depends on the observation through  a series of explanatory variables $x_{i0},...,x_{ik}$ associated with each $Y_{i}$ ($x_{i0}=1,$ $x_{ij}\in\mathbb{R}$, $i=1,...,n$,
$j=1,...,k$). We consider a $(k+1)$-dimensional vector of parameters $\boldsymbol{\beta}=\left(\beta_0,..,\beta_k\right)^T$, with $\beta_{i}\in\left(
-\infty,\infty\right),$ such that the binomial parameter, $\pi_{i} $, is linked to the linear predictor  via the
logit function, i.e.,
	$ \text{\textrm{logit}}\left(  \pi_{i}\right)  =\sum\limits_{j=0}^{k}\beta
	_{j}x_{ij}, \label{logit} $
where \textrm{logit}$(p)=\log(p/(1-p))$. This relation links the expectation of $Y_i$ (or equivalently the probability of belonging to the group labelled as 1) given the explanatory variables $\boldsymbol{x}_i$ as
\begin{equation}
	E[Y_i\lvert\boldsymbol{x}_i] = \pi_{i}=\pi(\boldsymbol{x}_{i}^{T}\boldsymbol{\beta})=\frac{e^{\boldsymbol{x}%
			_{i}^{T}\boldsymbol{\beta}}}{1+e^{\boldsymbol{x}_{i}^{T}\boldsymbol{\beta}}%
	},\text{ }i=1,...,n, \label{1.1}%
\end{equation}
where\ $\boldsymbol{x}_{i}^{T}=\left(  x_{i0},...,x_{ik}\right).$ 
The likelihood function for the logistic regression model is then given by
\begin{equation}
	\mathcal{L}\left(  \boldsymbol{\beta}\right)  =\prod\limits_{i=1}^{n}%
	\pi(\boldsymbol{x}_{i}^{T}\boldsymbol{\beta})^{y_{i}}\left(  1-\pi
	(\boldsymbol{x}_{i}^{T}\boldsymbol{\beta})\right)  ^{1-y_{i}}. \label{1.3}%
\end{equation}
where $y_{{1}},...,y_{n}$ denote the observed values of the random
variables $Y_{1},...,Y_{n}.$ Then, the MLE of $\boldsymbol{\beta}$, $\widehat{\boldsymbol{\beta}}$, is obtained minimizing the negative log-likelihood function over $\boldsymbol{\beta} \in \Theta$ with 
\[
\Theta=\left\{  \left(  \beta_{0},...,\beta_{k}\right)  ^{T}:\beta_{j}%
\in\left(  -\infty,\infty\right)  ,\text{ }j=0,...,k\right\}  =\mathbb{R}%
^{k+1}.
\]
In practice, if the number of observations, $n,$ is not large enough compared the parameter dimension $k$ (high dimensional set-up), the  classical logistic regression may cause over-fitting.
A useful technique to overcome this weakness is the regularization by adding a non-negative penalty $p_{\lambda_n}$ to the negative log-likelihood function (\ref{1.3}) where 
$\lambda_{n}>0$ is the
regularization parameter controlling the strength of shrinkage in the regression coefficients: the greater the value of $\lambda_{n}$ is, the greater weight will
be given to the penalty term and consequently the amount of shrinkage.
Without loss of generality, let us assume  that the covariates are standardized.
As a result, the intercept, $\beta_{0},$ is not penalized and the estimation of the vector $\boldsymbol{\beta}$ under a high-dimensional logistic regression model  is performed through the minimization of  the penalized likelihood-based objective function given by
\begin{equation}
	Q_{n,\lambda}\left(  \boldsymbol{\beta}\right)  =-\log\mathcal{L}\left(
	\boldsymbol{\beta}\right)  +\mathop{\textstyle \sum }_{j=1}^{k}p_{\lambda_{n}%
	}\left(  \left\vert \beta_{j}\right\vert \right)  . \label{1.3.1}%
\end{equation}
The regularized estimate of $\boldsymbol{\beta}$ is obtained by minimizing
(\ref{1.3.1}), i.e.,
\[
\widehat{\boldsymbol{\beta}}_{P}=\arg\min_{\boldsymbol{\beta\in}\Theta
}Q_{n,\lambda}\left(  \boldsymbol{\beta}\right)  .
\]
The value $\lambda_{n}$ depends on data and can be computed using a cross-validation method (\cite{James}), or a generalized information criterion (\cite{FanTang}) which consistently identifies the true model with asymptotic probability 1 in the ultra-high dimensional setting. From now on, we denote $\lambda = \lambda_n$ for the sake of simplicity. Note that $\lambda=0$ gives the MLE, whereas if $\lambda\rightarrow\infty$ all $\beta_{j}$ tend to $0$.

Among several penalties discussed in the literature, the general $\ell_q$-penalization is defined as\\
\noindent  $ \sum_{j=1}^{k} p_{\lambda_{n}} \left(\lvert \beta_{j}\lvert \right)  = \lambda \Vert \boldsymbol{\beta} \Vert_{q}^{q} = \lambda \sum_{j=1}^{k} \lvert\beta_{j}\lvert^{q}.$  The value $q=2$ yields $l_{2}$-regularized logistic regression, i.e., the Ridge procedure, which is particularly appropiate when there is multicollinearity between the explanatory variables (see \cite{du}, \cite{s} and \cite{lecessie}), but not useful in high-dimensional set-ups.
The value $q=1$ results in the $\ell_1$-penalization (LASSO) proposed by \cite{Tibshirani} in the context of linear regression. In this case, the penalty function continuously shrinks the coefficients toward zero, yielding a sparse subset of variables with nonzero regression coefficients; the larger the value of $\lambda$, greater the number of zero coefficients. \cite{Shevade} proposed sparse logistic regression based on the LASSO penalty and \cite{ca} investigated the same with Bayesian penalty. 
Some interesting results as well as applications of LASSO procedure in logistic regression can be seen in \cite{Fokianos}, \cite{Park}, \cite{Koh}, \cite{Yang}, \cite{plan}, \cite{Zhu} and \cite{Sun}. However, the LASSO penalty has been criticized for its biasedness, as it tends to select many noisy features (false positives) with high probability (\cite{huan}). To overcome the bias drawback, \cite{Zou} proposed the adaptive Lasso penalty  in the context of linear regression models. Precisely, the adaptive LASSO criterion for logistic regression considers the objective function,
\begin{equation}\label{adLASSO}
	Q_{n,\lambda}\left(  \boldsymbol{\beta}\right)  =-\log\mathcal{L}\left(
	\boldsymbol{\beta}\right)  +\lambda\mathop{\textstyle \sum }_{j=1}^{k}%
	\frac{\left\vert \beta_{j}\right\vert }{\vert \widetilde{\beta}_{j}\vert}, 
\end{equation}
where $\widetilde{\beta}_{j}$ is a consistent (initial) estimator of $\beta_{j}$ for all $j.$
Note that, for any fixed $\lambda$, the penalty for zero initial estimation of $\beta_{j}$ goes to infinity, while the weights for nonzero initials converge to a finite constant. Consequently, by allowing a relatively higher penalty for zero coefficients and lower for nonzero coefficients, the adaptive LASSO estimator reduces the estimation bias and improves variable selection accuracy.  Some interesting applications of adaptive LASSO can be seen in \cite{Algamal2015}, \cite{Algamal2019} and \cite{guo}. 

However, all the above methods are based on the negative log-likelihood function associated with the logistic regression model. The lack of robustness of the likelihood-based  loss is well known  and thus, leverage points or influence points can individually or jointly  have significant influence on the regularization procedures leading to erroneous conclusions. 
 \textcolor{black}{As an alternative, many robust penalized procedures are developed under high-dimensional linear regression models, with a few being also extended to the logistic regression set-ups. In particular, \cite{Wand2013} derived robust estimators in linear regression based on the exponential square loss, while \cite{Kawashima} applied $\gamma$-divergence for the same purpose. 
For the logistic regression model, \cite{bianco2021} studied general penalized M-estimators  in high-dimensional scenarios. }
\textcolor{black}{To achieve robustness against data contamination, we propose an alternative robust loss function based on the density power divergence (DPD) measure (\cite{gosh13}) along with the general adaptive loss. This idea  of using DPD based loss function in high-dimensional context  has been used previously in the context of high-dimensional linear regression models by \cite{ghosh20} and  \cite{gm} with non-concave and adaptive penalties, respectively. This work can be considered as an extension to the methods developed  therein  for robust modelling of  binary response data with high-dimensional logistic regression model.
}
We follow a similar idea in order to develop our robust DPD-based loss for the logistic regression model. In this respect, one may find the definitions of our adaptively weighted DPD-LASSO estimators (see Section \ref{sec2}) to be rather straightforward; but the derivations of their theoretical properties for the ultra-high dimensional logistic regression as well as its efficient  computation need non-trivial extensions of the existing literature as we will see throughout  the rest of the paper. These theoretical computations and numerical evaluation of the performance of the proposed estimator (over the existing procedures) are  our major contributions in this paper. In particular, we will show that our proposed estimators are indeed locally robust, through their influence function analysis (Section \ref{sec3}),  and enjoy oracle properties, i.e., consistently perform model selection and parameter estimation (Section \ref{sec4}). Additionally, we extend the IRLS algorithm, commonly used for obtaining the MLE in logistic regression, to the DPD-loss function in order to develop an efficient computational algorithm for our proposal (Section \ref{sec5}). 
Further,  we empirically compare our proposed estimators to some state-of-the-art robust and non-robust methods under the high-dimensional  logistic regression (Section \ref{sec6}). Results show significant  improvement in robustness of our proposal over all existing robust procedures under different contaminated scenarios, while keeping it competitive to the classical loglikelihood based methods in the absence of contamination. 

A major focus in cancer research is identifying genetic markers. The cancer disease is caused by abnormalities in the genetic material resulting from acquired mutations and epigenetic changes that influence the gene expression of a patient.
Microarray technology measures the expression level of genes of an individual and that genetic expression can be used in the diagnosis, through the classification of different types of cancerous genes leading to a cancer disease. Clinical diagnosis of cancer based on gene expression data pursue two goals, correctly diagnosing a patient and identifying the genes involved in this diagnosis. Moreover, microarray data are generally high dimensional data, having a large number of genes in comparison to the number of samples. Therefore, these two purposes can be achieved using regularized logistic regression models as a classifier, which also performs gene selection.  
In addition, it is well known that microarray datasets with many genes often contain outliers and several studies have pointed out that errors in labelling and gene expression measure are not uncommon, motivating the use of robust procedures.

We will apply our proposed estimation methods on four publicly available datasets which have been widely used to study different techniques in cancer classification, regarding three types of tumours; colon cancer, leukaemia and breast cancer. These four datasets have been flagged as being suspected of containing wrongly labelled samples or outlying observations, making robust methods particularly suitable for their analysis.  


%


\section{Adaptive LASSO based on the density power divergence loss \label{sec2}}
Firstly, we recall that the likelihood-based  loss (\ref{1.3.1}) under a logistic regression model  could also be justified from an information theoretic perspective by considering the famous Kullback-Leibler (KL) divergence measure between  the probability vectors $\widehat{\boldsymbol{p}}= \frac{1}{n} \left( y_{1},{1-y_{1}}, {y_{2}},{1-y_{2}},...,{y_{n}}, {1-y_{n}}\right)  ^{T}$ and
\[
\boldsymbol{p}\left(  \boldsymbol{\beta}\right)  =\frac{1}{n} \left(
\pi(\boldsymbol{x}_{1}^{T}\boldsymbol{\beta}), \left(  1-\pi(\boldsymbol{x}%
_{1}^{T}\boldsymbol{\beta})\right),...,\pi(\boldsymbol{x}_{n}%
^{T}\boldsymbol{\beta}), \left( 1-\pi(\boldsymbol{x}_{n}^{T}%
\boldsymbol{\beta})\right)\right)^{T}.
\]
Clearly, the vector $\widehat{\boldsymbol{p}}$ could be interpreted as the empirical estimator of $\boldsymbol{p}\left( \boldsymbol{\beta}\right).$ Mathematically, the KL divergence between $\widehat{\boldsymbol{p}}$ and $\boldsymbol{p}\left( \boldsymbol{\beta}\right)$ is given by
\begin{equation}
	d_{KL}\left(  \widehat{\boldsymbol{p}},\boldsymbol{p}\left(  \boldsymbol{\beta
	}\right)  \right)  =\sum\limits_{i=1}^{n}\left(  \frac{y_{i}}{n}\log
	\frac{y_{i}}{\pi(\boldsymbol{x}_{i}^{T}\boldsymbol{\beta})}+\frac{1-y_{i}}%
	{n}\log\frac{1-y_{i}}{1-\pi(\boldsymbol{x}_{i}^{T}\boldsymbol{\beta})}\right)
	. \label{1.5}%
\end{equation}
With some algebra, it is not difficult to establish that
\begin{equation}
	d_{KL}\left(\widehat{\boldsymbol{p}},\boldsymbol{p}\left(  \boldsymbol{\beta
	}\right)  \right)  =c-\frac{1}{n}\log\mathcal{L}\left(  \boldsymbol{\beta
	}\right)  , \label{1.6}%
\end{equation}
and therefore, the MLE of $\boldsymbol{\beta}$ can be defined by
\begin{equation}
	\widehat{\boldsymbol{\beta}}=\arg\min_{\boldsymbol{\beta\in}\Theta}%
	d_{KL}\left(  \widehat{\boldsymbol{p}},\boldsymbol{p}\left(  \boldsymbol{\beta
	}\right)  \right)  . \label{1.7}%
\end{equation}
Based on (\ref{1.7}), we might replace $d_{KL}$ by any divergence measure between the probability vectors $\widehat{\boldsymbol{p}}$ and $\boldsymbol{p}(\boldsymbol{\beta})$ 
in order to define a minimum distance estimator (MDE) for $\boldsymbol{\beta}$. In this paper we shall use the density power divergence (DPD) measure defined by \cite{basu1998}, since it produces MDEs with excellent robustness properties (see, e.g., Basu et al. (2011, 2013, 2015, 2016),
Ghosh et al. (2015, 2016)). 
The DPD between the probability vectors $\widehat{\boldsymbol{p}}$ and $\boldsymbol{p}\left(\boldsymbol{\beta}\right)$
is given by
\begin{align}
	d_{\alpha}\left( \widehat{\boldsymbol{p}},\boldsymbol{p}\left(\boldsymbol{\beta}\right)\right)  &  =\frac{1}{n^{1+\alpha}}\sum \limits_{i=1}^{n}\left\{ \left(  \pi^{1+\alpha}(\boldsymbol{x}_{i}^{T}\boldsymbol{\beta})+\left(  1-\pi(\boldsymbol{x}_{i}^{T}\boldsymbol{\beta})\right)^{1+\alpha}\right)   \right. \nonumber\\
	&  -\left( 1+\frac{1}{\alpha}\right) \left(  y_{i}\pi^{\alpha}(\boldsymbol{x}_{i}^{T}\boldsymbol{\beta
	}) +(1-y_{i})\left(1-\pi(\boldsymbol{x}_{i}^{T}\boldsymbol{\beta})\right)
	^{\alpha}\right) \nonumber\\
	& \left. + \frac{1}{\alpha}\left( y_{i}^{\alpha+1}+(1-y_{i}%
	)^{\alpha+1}\right)  \right\} \nonumber \\
	&= \frac{1}{n^{1+\alpha}}\sum \limits_{i=1}^{n} \rho_{\alpha}(\boldsymbol{x}_i^{T}\boldsymbol{\beta}, y_i) .\label{dalpha}
\end{align}
for $\alpha>0$ (see \cite{basu2017}) where
\begin{equation*} \label{rho}
	\begin{aligned}
		\rho_{\alpha}(\boldsymbol{x}^{T}\boldsymbol{\beta}, y) = &   \frac{1}{n^\alpha} \left[\left(\pi^{1+\alpha} +\left(  1-\pi\right)^{1+\alpha} \right)  
		- \left(  1+\frac{1}{\alpha}\right)  \left(  y\pi^{\alpha}+(1-y)\left(1-\pi \right)^{\alpha}\right) \right. \\
		& \left. +\frac{1}{\alpha}\left( y^{\alpha+1}+(1-y)^{\alpha+1}\right)\right].
	\end{aligned}
\end{equation*} 
Note that the third term does not depend on $\boldsymbol{\beta}$ and therefore it can be excluded in the minimization. The value $\alpha=0$ is  derived from the general formula by taking its continuous
limit as $\alpha$ tends to $0,$ which indeed yields
\[ d_{0}\left(  \widehat{\boldsymbol{p}},\boldsymbol{p}\left(\boldsymbol{\beta
}\right)  \right)  =\lim_{\alpha\rightarrow0}d_{\alpha}\left( \widehat{\boldsymbol{p}},\boldsymbol{p}\left(  \boldsymbol{\beta}\right)
\right)  =d_{KL}\left(  \widehat{\boldsymbol{p}},\boldsymbol{p}\left(
\boldsymbol{\beta}\right)  \right)  . \]
Hence, the minimum density power divergence estimator (MDPDE), $\widehat{\boldsymbol{\beta}}_{\alpha}$, for the parameter $\boldsymbol{\beta
}$, in the logistic regression model is defined as the minimizer of $d_\alpha(\widehat{\boldsymbol{p}},\boldsymbol{p}\left(\boldsymbol{\beta}\right))$ in (\ref{dalpha}), which coincides with the MLE at $\alpha=0$ and provides a robust generalization of it at $\alpha>0$. 

Taking derivatives respect to $\boldsymbol{\beta}$ we obtain the estimating equations of the MDPDE at $\beta>0$ as given by
\begin{equation} \label{1.61}
	\sum\limits_{i=1}^{n}\boldsymbol{\Psi}_{\alpha}\left(  \boldsymbol{x}_{i},y_{i},\boldsymbol{\beta}\right)\boldsymbol{x}%
	_{i}  =\boldsymbol{0}_{k+1}, 
\end{equation}
with
\begin{equation}\label{Psi}
	\boldsymbol{\Psi}_{\alpha}\left(  \boldsymbol{x}_{i},y_{i},\boldsymbol{\beta
	}\right)  =(e^{\alpha\boldsymbol{x}_{i}^{T}\boldsymbol{\beta}}%
	+e^{\boldsymbol{x}_{i}^{T}\boldsymbol{\beta}})\frac{e^{\boldsymbol{x}_{i}%
			^{T}\boldsymbol{\beta}}-y_{i}(1+e^{\boldsymbol{x}_{i}^{T}\boldsymbol{\beta}}%
		)}{(1+e^{\boldsymbol{x}_{i}^{T}\boldsymbol{\beta}})^{\alpha+2}},
\end{equation}
which reconfirms that the MDPDE in logistic regression models is an M-estimator but with a model-dependent $\Psi$-function (\ref{Psi}). 
For the value $\alpha=0$ in (\ref{1.61}) we get the estimating equations for the MLE which are 
\[ \sum\limits_{i=1}^{n}\left(  \pi(\boldsymbol{x}_{i}^{T}\boldsymbol{\beta
})-y_{i}\right)  \boldsymbol{x}_{i}=\boldsymbol{0}_{k+1}. \]

In this paper, utilizing  the idea of adaptive LASSO penalty discussed in Section \ref{sec1}, we propose to minimize the regularized objective function with a general adaptively weighted LASSO penalty as given by 
\begin{equation}
	Q_{n,\alpha,\lambda}\left(  \boldsymbol{\beta}\right)  =d_\alpha(\widehat{\boldsymbol{p}},\boldsymbol{p}\left(\boldsymbol{\beta}\right)) +\lambda\mathop{\textstyle \sum }_{j=1}^{k}w(\lvert\widetilde{\beta}_{j}\lvert)\lvert \beta_{j}\lvert , \label{1.101}
\end{equation}
where $w$ is a suitable weight function, $\lambda$ is the usual regularization parameter and $ \widetilde{\boldsymbol{\beta}} =  \left( \widetilde{\beta}_{j}\right) _{j=1,..,k}$ is any consistent estimator of $\boldsymbol{\beta}$, referred to as the initial estimator. 
We will denote the estimator obtained minimizing by $Q_{n,\alpha,\lambda}\left( \boldsymbol{\beta}\right) $ in (\ref{1.101}) by $\widehat{\boldsymbol{\beta}}_{\lambda,\alpha}$ and call it an adaptive weighted DPD-LASSO estimator (AW-DPD-LASSO) of parameter $\boldsymbol{\beta}$ under the high-dimensional logistic regression model.  When the weight function is $w = 1$, we obtain the DPD-based regularized with standard LASSO estimator (which we refer to  as the DPD-LASSO). When the conventional adaptive LASSO penalty, obtained by using   the hard-thresholding weight function  $w(s )=\frac{1 }{s }I\left( s \neq0\right)$ is used the corresponding estimator will be  referred to as the adaptive DPD-LASSO (Ad-DPD-LASSO). 
\begin{remark}
	\cite{ghoshbasu2016} derived the expression of the DPD loss for generalized regression models, where $Y_i$ follows the general exponential family of distributions. For independent but not identically distributed observations, they minimized the average divergence between the data points and the models. 
	Using Bernoulli distributions, they obtained the DPD loss 
	\begin{equation} \label{2.14}
		\frac{1}{n}\sum_{i=1}^n \left\{(1-\pi(\boldsymbol{x}_{i}^{T}\boldsymbol{\beta}))^{\alpha+1}+\pi^{\alpha+1}(\boldsymbol{x}_{i}^{T}\boldsymbol{\beta}) - \frac{\alpha+1}{\alpha} \pi^{\alpha y_i}(\boldsymbol{x}_{i}^{T}\boldsymbol{\beta}) (1-\pi(\boldsymbol{x}_{i}^{T}\boldsymbol{\beta}))^{\alpha(1-y_i)}\right\}
	\end{equation}  
	Note that our DPD-based loss function given in  (\ref{dalpha}) differs from the above expression in  (\ref{2.14}) only on the scale $\frac{1}{n^\alpha}$ since $Y_i$ can take only values 0 or 1. Therefore, both expressions achieve the minimum at the same point without regularization.
\end{remark}

\begin{remark} \label{remarkscad}
	\textcolor{black}{
	Adaptively weighted penalties can also provide computationally flexible approximation to any general non-concave penalty functions, $p_{\lambda_n}(.)$, using appropriate first order Taylor expansion series around an initial estimator, when the weight function is chosen as $w(s) = p_{\lambda_n}'(s)$; see \cite{fan1} and \cite{fan2014} for details.} In particular, the weight function approximating the popular SCAD penalty is given by 
	\begin{eqnarray}
		w(\lvert\beta_j\lvert)  =I(\lvert\beta_j\lvert\leq \lambda_n) + \frac{(a\lambda_n - \lvert\beta_j\lvert)_{+}}{(a-1)\lambda_n}I(\lvert\beta_j\lvert>\lambda_n), 
		\label{EQ:SCAD}
	\end{eqnarray}
	with $a>2$ being a tuning constant, suggested as $a=3.7$ on an empirical basis. 
\end{remark}

\section{Robustness: Influence Function Analysis} \label{sec3}

We study the local robustness of the AW-DPD-LASSO  through its influence function (IF), which quantifies the impact that entails a infinitesimal contamination in the sample on parameter estimation. An estimator is said locally robust if it associated IF is bounded.
We consider $\left( Y_{1},\boldsymbol{X}_{1}\right),...,\left(  Y_{n},\boldsymbol{X}_{n}\right), $ a random sample from the random vector $\left( Y,\boldsymbol{X}\right)$ with true distribution function $G\left(  y,\boldsymbol{x}\right).$  We assume that $\boldsymbol{X}_{1},...,\boldsymbol{X}_{n}$ is a random sample from a random variable $\boldsymbol{X}$ with marginal distribution function $H(\boldsymbol{x})$. The statistical functional $\boldsymbol{T}_{\lambda,\alpha}^{\boldsymbol{\beta}}(G)$ corresponding to the AW-DPD-LASSO estimator $\widehat{\boldsymbol{\beta}}_{\lambda,\alpha}$ is defined as the minimizer of

\begin{equation}
	Q_{\alpha,\lambda}\left(  \boldsymbol{\beta}\right) = \int \frac{1}{n^\alpha}\rho_\alpha(\boldsymbol{x}^T\boldsymbol{\beta}, y)  dG\left(  y,\boldsymbol{x}\right)  +\lambda \sum\limits_{j=1}^{k}w\left(\lvert U_{j}(G) \lvert \right) \lvert\beta_{j}\lvert , \label{2.1}
\end{equation}
with respect to $\boldsymbol{\beta},$ where $\boldsymbol{U(G)} = \left(U_{1}(G),...,U_{k}(G)\right)$ is the statistical functional corresponding to the initial estimator $\widetilde{\boldsymbol{\beta}}.$
Note that the objective function $Q_{\alpha,\lambda}\left(\boldsymbol{\beta}\right)$ in (\ref{2.1}) coincides with the empirical objective function (\ref{1.101}) when $G$ is substituted by the empirical distribution function $G_{n}$, as $\boldsymbol{U}(G_{n}) = \widetilde{\boldsymbol{\beta}},$ and hence, $\boldsymbol{T}_{\lambda,\alpha}^{\boldsymbol{\beta}}(G_{n}) = \widehat{\boldsymbol{\beta}}_{\lambda,\alpha}.$  
As mentioned in Section \ref{sec2}, the MDPDE belongs to the family of M-estimators, so we can apply the extended IF theory developed in \cite{Avella-Medina:2017} for our proposed MDPDE, with their $L\left(\boldsymbol{\theta},\boldsymbol{Z}\right)$ being our $L_{\alpha}\left(\left(  y,\boldsymbol{x}\right);\boldsymbol{\beta}\right)$. They justified that the IF of a M-estimator of the form (\ref{2.1}) can be attainted as
\begin{equation} \label{limIF}
	\text{IF}\left((y_t, \boldsymbol{x}_t), \boldsymbol{T}_{\lambda,\alpha}^{\boldsymbol{\beta}}(G),G\right) = \lim_{m \rightarrow \infty} \text{IF}_{p_m}\left((y_t, \boldsymbol{x}_t), \boldsymbol{T}_{\lambda,\alpha,m}^{\boldsymbol{\beta}}(G), G\right),
\end{equation}
where $\text{IF}_{p_m}$ are the influence functions of estimators penalized with continuous and infinitely differentiably functions in both arguments $p_m(s, t(G))$, for all $m = 1,2,...,$ 
and $p_m(s, t(G))$ converges in the Sobolev space to our adaptive weighted penalty $w\left(|t(G)| \right) |s|$ as $m\rightarrow \infty$.
\cite{Avella-Medina:2017} proved that the expression of $\text{IF}\left((y_t, \boldsymbol{x}_t), \boldsymbol{T}_{\lambda,\alpha}^{\boldsymbol{\beta}}(G), G\right)$ does not depend on the election of the sequence $p_m(s, t(G))$, and futher, if the  parameter space is compact and satisfies certain regularity conditions involving the continuity in $\boldsymbol{\beta}$ of some relevant functions and the invertibility  of the Hessian of the corresponding objective function at $\boldsymbol{\beta}$ for all $m$, the $\text{IF}$ exists for all $(y_t,\boldsymbol{x}_t).$

Let us consider $s$ to be the model size of true data-generating model  and  let us denote 
$\boldsymbol{\beta}_1,$ $\boldsymbol{x}_{1}$ and $\boldsymbol{x}_{t,1}$ the $s$-vectors containing the first $s$ elements of $\boldsymbol{\beta},$ $\boldsymbol{x}$ and $\boldsymbol{x}_t$,  and the corresponding functional $\boldsymbol{T}_{\lambda,\alpha}^{\boldsymbol{\beta}}(G) = \left( \boldsymbol{T}_{\lambda,\alpha}^{\boldsymbol{\beta}_1}(G), \boldsymbol{T}_{\lambda,\alpha}^{\boldsymbol{\beta}_{k-s}}(G)\right).$ 
Then, the IF of our AW-DPD-LASSO estimator, $\boldsymbol{T}_{\lambda,\alpha}^{\boldsymbol{\beta}},$ is presented in the following theorem.

\begin{figure}[htb]
	\centering
	\caption{$\ell_2$-norm of the Ad-DPD-LASSO Influence Function, with fixed parameters $\boldsymbol{\beta} = (3,2),\lambda = 0.1$ and $n=100.$ The IF for the value $\alpha=0$ is rescaled $(\times 10^{-2}).$ }
	\label{figure}
	\includegraphics[height=6cm, width= 9cm]{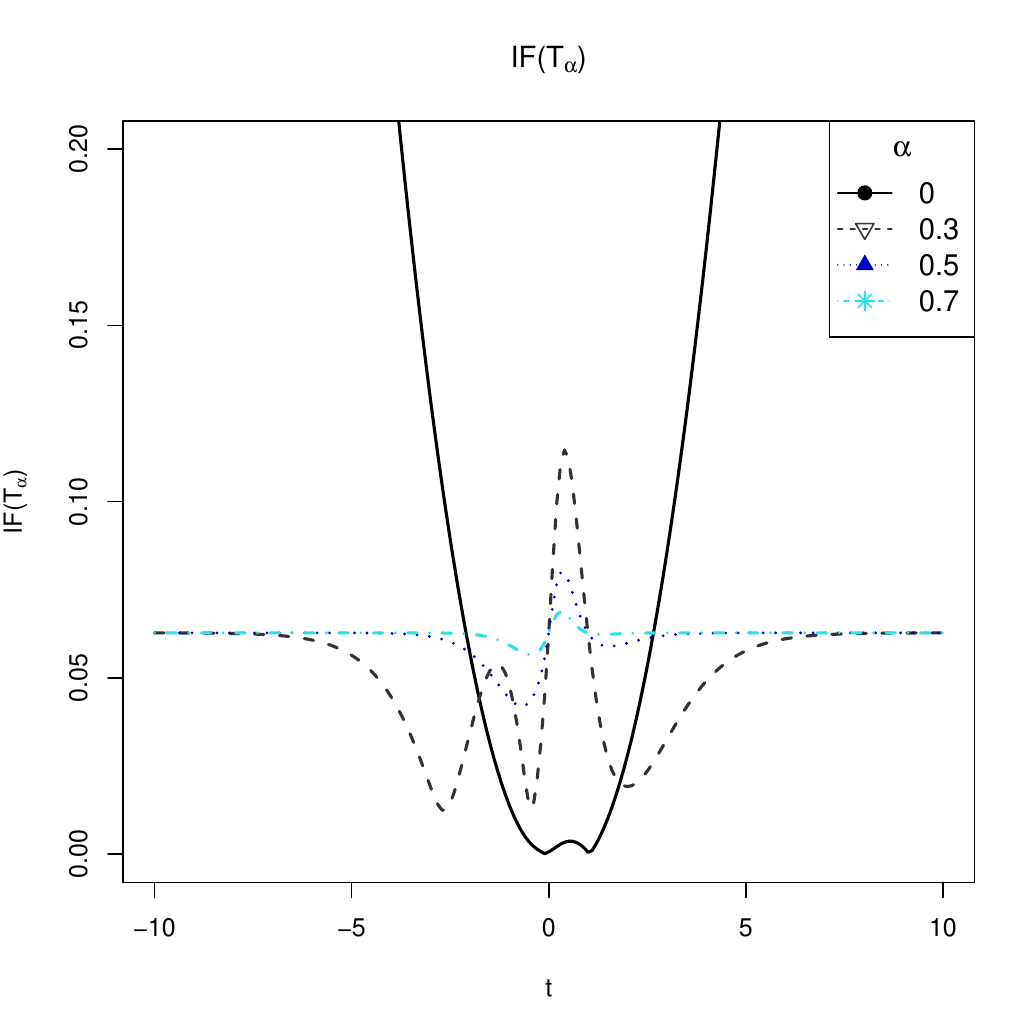}
\end{figure}

\begin{theorem} \label{thmIF}
	Consider the previous set-up and the true parameter value $\boldsymbol{\beta}^g = \boldsymbol{T}_{\lambda,\alpha}^{\boldsymbol{\beta}}(G),$ where $G$ is the true distribution underlying the data, $\boldsymbol{\beta}^g$ is sparse with only $s < k$ nonzero components and $k \gg n.$ Without loss of generality, assume $\boldsymbol{\beta}^g = \left(\boldsymbol{\beta}^{g T}_1, \boldsymbol{\beta}^{g T}_{k-s}\right)^T,$ where $\boldsymbol{\beta}^g_1$ contains all $s-$nonzero elements of 
	$\boldsymbol{\beta}^g.$ Define 
	$\boldsymbol{P}^\ast(\boldsymbol{\beta}, \boldsymbol{U}(G)) $ a $k$-vector having $j$-th element as $ w(|U_j(G)|)\operatorname{sign}(\beta_j)$
	and
	$\boldsymbol{P}^{\ast(2)}(\boldsymbol{\beta},\boldsymbol{U}(G))$ as the $s \times s$ diagonal matrix with $j$-th entry being  $ w'(U_j(G)) \operatorname{sign}(U_j(G)\beta_j)$ for $j=1,2,\ldots, s$ and 
	$ \boldsymbol{S}_\alpha(G,\boldsymbol{\beta}^g) $ the $s \times s$ submatrix containing the first $s$ columns and rows of $J_\alpha(G,\boldsymbol{\beta}^g),$ with
	\begin{equation}
		\begin{aligned} \label{matrixJast}
			J_\alpha(G,\boldsymbol{\beta}) = \frac{\alpha+1}{n^\alpha}\mathbb{E}_{\boldsymbol{X}} \bigg[ \left( 1 + e^{\boldsymbol{x}^T\boldsymbol{\beta}}\right)^{-(\alpha+3)}
			\bigg[ e^{(\alpha+1)\boldsymbol{x}^T\boldsymbol{\beta}} - e^{2\boldsymbol{x}^T\boldsymbol{\beta}}
			\bigg] \boldsymbol{x}\boldsymbol{x}^T\bigg]. 
		\end{aligned}
	\end{equation}
	Then, whenever the associated quantities exist finitely, the $\text{IF}((y_t,\boldsymbol{x}_t), \boldsymbol{T}_{\lambda,\alpha}^{\boldsymbol{\beta}_{k-s}}, G)$ is identically zero and the IF of $\boldsymbol{T}_{\lambda,\alpha}^{\boldsymbol{\beta}_{1}}$ is given by
	\begin{equation} \label{IF}
		\begin{aligned}
			\text{IF}((y_t,\boldsymbol{x}_t), \boldsymbol{T}_{\lambda,\alpha}^{\boldsymbol{\beta}_{1}}, G) = & - \boldsymbol{S}_\alpha(G,\boldsymbol{\beta}^g)^{-1}\bigg[ 
			\frac{1+\alpha}{n^{\alpha}} \boldsymbol{\Psi}_\alpha \left(  \left(  y_t,\boldsymbol{x}_{t,1}\right) ;\boldsymbol{\beta}_1\right)\boldsymbol{x}_{t,1} \\
			& + \lambda \boldsymbol{P}^{\ast }_{\lambda}(\boldsymbol{\beta}_1,\boldsymbol{U}(G)) +  \lambda\boldsymbol{P}^{\ast(2)}(\boldsymbol{\beta}_{1},\boldsymbol{U}(G))\text{IF}\left((y_t, \boldsymbol{x}_{t,1}), \boldsymbol{U}_1, G\right) \bigg]
		\end{aligned}
	\end{equation}
	where $\boldsymbol{U}_1$ denotes the first $s$ elements of $\boldsymbol{U}(G).$ 
\end{theorem}
\textcolor{black}{We have provided the details of  all calculations for Theorem \ref{thmIF} in Section 1 of the Supplementary material  available Online. }
If we consider the fixed design matrix set-up, the vector $\boldsymbol{x}$ is non-stochastic and the expectation in (\ref{matrixJast}) should be ignored.
Using Theorem \ref{thmIF} we can compute the expression of the $\text{IF}$ of $\boldsymbol{T}_{\lambda,\alpha}^{\boldsymbol{\beta}_{1}}$ at the true model distribution $G.$ Note that $\boldsymbol{\beta}_1^g =\boldsymbol{\beta}_1$ at the true model $G,$ and $\boldsymbol{U}(G) = \boldsymbol{\beta}$ because the initial estimator is consistent. It is clear that the boundness of $\text{IF}((y_t,\boldsymbol{x}_{t,1}), \boldsymbol{T}_{\lambda,\alpha}^{\boldsymbol{\beta}_{1}}, G)$ depends only on $\boldsymbol{\Psi}_\alpha \left(  \left(  y_t,\boldsymbol{x}_{t,1}\right) ;\boldsymbol{\beta}_1\right)\boldsymbol{x}_{t,1}$ and $\text{IF}\left((y_t, \boldsymbol{x}_{t,1}), \boldsymbol{U}_1, G\right)$. The second function is bounded only if the initial estimator $\tilde{\boldsymbol{\beta}}$ is robust; thus a robust initial estimator is needed to achieve robustness for our proposed estimator AW-DPD-LASSO.
Finally, note that 
$\boldsymbol{\Psi}_\alpha \left(  \left(  y_t,\boldsymbol{x}_{t,1}\right) ;\boldsymbol{\beta}_1\right) \boldsymbol{x}_{t,1}$ as defined in (\ref{Psi}) is bounded either or both of the contamination point $(y_t, \boldsymbol{x}_t).$
Figure \ref{figure} shows the $\ell_2$-norm of the influence function of Ad-DPD-LASSO with identical contamination in all directions, $\text{IF}(t(1,\boldsymbol{1}), \boldsymbol{T}_{\lambda,\alpha}^{\boldsymbol{\beta}_{1}}, G),$ $t\in \mathbb{R},$ for different values of the parameter $\alpha$ and using the DPD-LASSO as initial estimator. A particular choice of true regression coefficients $\boldsymbol{\beta} = (3,2),$ parameter $\lambda = 0.1$ and sample size $n=100$ was made.  
$\text{IF}((y_t,\boldsymbol{x}_{t,1}), \widehat{\boldsymbol{\beta}}_{\lambda,\alpha,1}^{\text{\tiny{LASSO}}}, G)$ can be easily obtained from expression (\ref{IF}) by substituting $w(s) \equiv 1$. As shown, the IF of the Ad-DPD-LASSO is bounded for positives values of $\alpha$, whereas remains unbounded for $\alpha=0.$

\section{Oracle properties}\label{sec4}
We now study the asymptotic properties of the proposed AW-DPD-LASSO estimators under the ultra-high dimensional set-up with non polynomial order, i.e., when $\log(k) = \mathcal{O}(n^\zeta),$ for some $\zeta \in (0,1)$,  by establishing its oracle properties under certain necessary conditions. 
For simplicity we assume that the design matrix $\mathbb{X}$ is fixed with each column being standardized to have $\ell_1$-norm $\sqrt{n}$,  which also implies that the $\ell_2$-norm of each column is also bounded by $\sqrt{n}$. \textcolor{black}{ Note that the second bound on $\ell_2$-norm is more standard and weaker than our assumed bound on $\ell_1$-norm, but this little stricter bound is needed to get the desired rates in our theoretical derivations with the DPD loss function as was also seen previously in \cite{gm}.}

We first introduce some useful notation. Given any $\mathcal{S} \subset \{1,2...,p\}$ and any $p$-vector $\boldsymbol{v}=(v_1, \ldots, v_p)^T$, 
we will denote $\mathcal{S}^c = \{1, 2, \ldots, p\} \setminus \mathcal{S}$,  $\boldsymbol{v}_S=(v_j : j\in S)$ whereas $Supp(\boldsymbol{v}) = \{ j : v_j \neq 0 \}$.
Recall that the true model is assumed to be sparse, and so if we denote $\mathcal{S}_0$ to be the subset of non zero elements of the true regression vector, $\boldsymbol{\beta}_0$, then the cardinality of $\mathcal{S}_0$ is $s \ll n$. Without loss of generality, we can assume that the first $s$ elements of $\boldsymbol{\beta}_{0}$ are non-zero. In particular, let us allow that $s=s_n=o(n)$ slowly diverges with the sample size $n$.
Further, let $\mathbb{X}_{\mathcal{S}}$ consist of the $j$-th column of $\mathbb{X}$ for all  $j\in \mathcal{S}$ for any $\mathcal{S}$, 
and put $\mathbb{X}_1=\mathbb{X}_{\mathcal{S}_0}$ and $\mathbb{X}_2=\mathbb{X}_{\mathcal{S}_0^c}$ 
(so that $\mathbb{X}=[\mathbb{X}_1 , \mathbb{X}_2]$);
the corresponding partition of the $i$-th row of $\mathbb{X}$ would be denoted by $\boldsymbol{x}_i = (\boldsymbol{x}_{1i}^T, \boldsymbol{x}_{2i}^T)^T$. 
The gradient and Hessian matrix of the DPD loss is given by
$
\frac{\partial d_\alpha(\widehat{\boldsymbol{p}},\boldsymbol{p}\left(\boldsymbol{\beta}\right))}{\partial \boldsymbol{\beta} } = \frac{1}{n}\mathbb{X}^T\boldsymbol{H}^{(1)}_{\alpha}(\boldsymbol{\beta}),$ and 
$
\frac{\partial^2 d_\alpha(\widehat{\boldsymbol{p}},\boldsymbol{p}\left(\boldsymbol{\beta}\right))}{\partial \boldsymbol{\beta} \boldsymbol{\beta}^T} = \frac{1}{n} \mathbb{X}^T \boldsymbol{H}^{(2)}_{\alpha}(\boldsymbol{\beta})\mathbb{X},
$
with $\boldsymbol{H}^{(1)}_{\alpha}(\boldsymbol{\beta})$ being a $(k+1)$-dimensional vector with $i$-component $\boldsymbol{H}^{(1)}_{\alpha}(\boldsymbol{\beta})_i = \frac{(\alpha+1)}{n^\alpha} \boldsymbol{\Psi}_{\alpha}\left(  \boldsymbol{x}_{i},y_{i},\boldsymbol{\beta
}\right)$, where $ \boldsymbol{\Psi}_{\alpha}\left( \boldsymbol{x}_{i},y_{i},\boldsymbol{\beta}\right)$ is as defined in (\ref{Psi})
and $\boldsymbol{H}^{(2)}_{\alpha}(\boldsymbol{\beta})$ is a diagonal matrix with $i$-diagonal entries
\begin{equation}
	\begin{aligned} \label{hessian}
		&\boldsymbol{H}^{(2)}_{\alpha}(\boldsymbol{\beta})_{ii} \\
		& = 
		\frac{(\alpha+1)}{n^\alpha} \sum_{i=1}^n \bigg[ \pi_i^{\alpha+1}(1-\pi_i)\big((1+\alpha)-(2+\alpha)\pi_i\big)  + (1-\pi_i)^\alpha\pi^2\big(2-(2+\alpha)\pi_i\big)\\
		&  \hspace{0.5cm} -y \bigg[\pi_i^\alpha(1-\pi_i)\big(\alpha-(1+\alpha)\pi_i\big) + \pi_i(1-\pi_i)^\alpha\big(1-(\alpha+1)\pi_i\big)\bigg] \bigg].
	\end{aligned}
\end{equation}

Since AW-DPD-LASSO estimators coincide with the negative loglikelihood-based weighted adaptive LASSO at $\alpha=0$ we only derive the results for $\alpha >0$. 
We first establish the oracle property for the AW-DPD-LASSO with the assistance of the oracle information on the location of signal covariates, $\mathcal{S}_0,$ and then  extend the result to the cases of unknown  $\mathcal{S}_0$, and the $s$ first elements of the estimated regression vector are consistent estimators of $\boldsymbol{\beta}_{\mathcal{S}_0}.$ 
We start with the following assumptions, where all expectations are taken with respect to the conditional distribution $Y_i \lvert \boldsymbol{X}_i = \boldsymbol{x}_i$ (which is also the case for the rest of the paper).

\begin{enumerate}
	\item[\textbf{(A1)}] \label{a1} The function $\rho_{\alpha}(\cdot, y_i)$ is Lipschitz with Lipschitz constant $L$ for all $i=1,2,...,n.$ 
	\item[\textbf{(A2)}] \label{a2} The eigenvalues of the matrix 
	$n^{-1}(\boldsymbol{X}_1^T\boldsymbol{X}_1)$ are bounded below and above by positive constants $c_0$ and $c_0^{-1}$, respectively.
	Also $ \kappa_n := \max_{i,j}|x_{ij}| = o(n^{1/2}s^{-1}).$
	\item[\textbf{(A3)}] \label{a3} The diagonal elements of $E[\textcolor{black}{\boldsymbol{H}_{\alpha}^{(1)}(\boldsymbol{\beta}_0)}]$ are all finite and 
	bounded from below by a constant $c_1>0$.
	\item[\textbf{(A4)}] \label{a4} Expectation of all third order partial derivatives of  $\rho_{\alpha}(\boldsymbol{x}_i^T\boldsymbol{\beta}, y_i)$, $i=1, \ldots, n$, 
	with respect to the components of $\boldsymbol{\beta}_{S_0}$ are uniformly bounded in a neighborhood of $\boldsymbol{\beta}_{01}$. 
\end{enumerate}

The first two Conditions are common in the literature of robust high-dimensional inference; see, e.g., \cite{gm} and \cite{fan2014}. In particular, Lipschitz property of a density or a loss function is commonly assumed to achieve robust inference. 
Next two assumptions (A3)-(A4) on the DPD-loss function are are taken from the literature of classical minimum divergence inference, which hold for common (bounded) design matrices. 
Let us first restrict to the cases of  the fixed weights $\boldsymbol{w} = (w_1, \ldots, w_p)$ with $\boldsymbol{w}_0=\boldsymbol{w}_{\mathcal{S}_0}$, $\boldsymbol{w}_1 =\boldsymbol{w}_{\mathcal{S}_0^c}$, and define $\delta_n = \sqrt{{s(\log n)}/{n}} + {\lambda_n}\parallel\boldsymbol{w}_0\parallel_2$. 
Let us denote $\widehat{\boldsymbol{\beta}}^o$ the oracle estimator that minimizes the objective function (\ref{1.101}) over the restricted (to the true model) parameter space, $\left\{ \boldsymbol{\beta}=(\boldsymbol{\beta}_1^T, \boldsymbol{\beta}_2^T)^T\in \mathbb{R}^p : \boldsymbol{\beta}_2 = \boldsymbol{0}_{p-s} \right\}
\equiv \mathbb{R}^s\times \{0\}^{p-s}.$
Then, we obtain the $\ell_2$-consistency of the oracle estimator  $\widehat{\boldsymbol{\beta}}^o$ as presented in the following theorem; the proof is given in the Supplementary Material.

\begin{theorem}\label{THM:FixW_Oracle}
	Consider the cases of the fixed weights $\boldsymbol{w}=(w_1, \ldots, w_k)$ which satisfy the condition  $\lambda_n\parallel\boldsymbol{w}_0\parallel_2\sqrt{s}\kappa_n \rightarrow 0$ with $\boldsymbol{w}_0 = w_{S_0}$, and let Assumptions (A1)-(A4) hold.
	Then, given any  constant $C_1>0$ and $\delta_n = \sqrt{{s(\log n)}/{n}} + {\lambda_n}\parallel\boldsymbol{w}_0\parallel_2$, 
	there exists some $c>0$ such that the oracle estimator $\boldsymbol{\beta}^o = (\boldsymbol{\beta}^o_1, \boldsymbol{0})$ satisfies 
	\begin{eqnarray}
		P\left(\parallel\widehat{\boldsymbol{\beta}}_1^o - \boldsymbol{\beta}_{10}\parallel_2 \leq C_1\delta_n \right)\geq 1 - n^{-cs}.
	\end{eqnarray}
	Further, if $\delta_n^{-1}\min\limits_{1\leq j \leq s}\lvert \beta_{j0} \lvert \rightarrow \infty$, 
	then the sign of each component of $\widehat{\boldsymbol{\beta}}_1^o$ matches with that of the true $\boldsymbol{\beta}_{10}$.
\end{theorem}


Note that Theorem \ref{THM:FixW_Oracle} indicates the oracle rate to be within a factor of log $n$ and bias due to penalization. This result can be further extended to show that the AW-DPD-LASSO estimator enjoys the oracle property even when the oracle information is not known, for a convenient choice of the weight vector. Since the regularized estimator depends on the full design matrix, we need to make additional assumptions controlling the correlation between the significant and insignificant  explanatory variables allowed to have oracle model selection consistency. Following \cite{gm} and \cite{fan2014}, we assume

\begin{itemize}
	\item[\textbf{(A5)}]  $ \parallel n^{-1}(\boldsymbol{X}_2^TE[\boldsymbol{H}_{\alpha}^{(2)}(\boldsymbol{\beta}_0)]\boldsymbol{X}_1)\parallel_{2, \infty}
	< \frac{\lambda_n\min(\lvert\boldsymbol{w}_1\lvert)}{2C_1\delta_n} $, 
	for some constant $C_1 >0$ and $\delta_n$ as in Theorem \ref{THM:FixW_Oracle}, where we denote 
	$\parallel\boldsymbol{A}\parallel_{2, \infty} =\sup\limits_{\boldsymbol{x}\in\mathbb{R}^q \setminus\{0\}} 
	\frac{\parallel\boldsymbol{A}\boldsymbol{x}\parallel_{\infty}}{\parallel\boldsymbol{x}\parallel_2}$
	for any $p\times q$ matrix $\boldsymbol{A}$ and $\min(\lvert\boldsymbol{w}_1\lvert)=\min\limits_{j>s}\lvert w_j\lvert$.
	
\end{itemize}
The following theorem then shows  that the (unrestricted) oracle estimator is indeed an asymptotic global minimizer of our objective function over the whole parameter space with probability tending to one.
\begin{theorem}
	\label{THM:FixW_Main}
	Consider the cases of the fixed weights.  Suppose that Assumptions (A1)--(A5) hold with $\lambda_n > 2\sqrt{(c+1)\log p / n}$ 
	and $\min(\lvert\boldsymbol{w}_1\lvert) >c_3$ for some constant $c, c_3>0$, and
	$$
	\lambda_n\parallel\boldsymbol{w}_0\parallel_2\kappa_n\max\{\sqrt{s}, \parallel\boldsymbol{w}_0\parallel_2\} \rightarrow 0,
	~~~~
	\delta_ns^{3/2}\kappa_n^2(\log_2 n)^2 = o(n\lambda_n^2).
	$$
	Then, with probability at least $1 - O(n^{-cs})$, 
	there exists a global minimizer $\widehat{\boldsymbol{\beta}} = \left((\widehat{\boldsymbol{\beta}}_1^o)^T, \widehat{\boldsymbol{\beta}}_2^T\right)^T$ 
	of the AW-DPD-LASSO objective function $Q_{n,\gamma,\lambda }(\boldsymbol{\beta })$ in (\ref{1.101}) such that
	$$
	\left\lvert\left\lvert\widehat{\boldsymbol{\beta}}_1^o - \boldsymbol{\beta}_{10}\right\lvert\right\lvert_2 \leq C_1\delta_n,
	~~~~\mbox{and }~~~~~
	\widehat{\boldsymbol{\beta}}_2 = \boldsymbol{0}_{p-s}.
	$$
\end{theorem}

We next establish the asymptotic distribution of the AW-DPD-LASSO. For this purpose, we need to make additional assumptions to put further restrictions to ensure that the bias term due to penalization does not diverge.  
Let us define $\boldsymbol{V}_n=\left[\boldsymbol{X}_1^TE[\boldsymbol{H}_{\alpha}^{(2)}(\boldsymbol{\beta}_{0})]\boldsymbol{X}_1\right]^{-1/2}$, $\boldsymbol{Z}_n=\boldsymbol{X}_1\boldsymbol{V}_n$ and $\boldsymbol{\Omega}_n = Var\left[\boldsymbol{H}_{\alpha}^{(1)}(\boldsymbol{\beta}_{0})\right]$, and consider the following assumption. 
\begin{enumerate}
	\item[(A6)]  $\boldsymbol{Z}_n^T\boldsymbol{\Omega}_n\boldsymbol{Z}_n$ is positive definite, $\lambda_n \parallel\boldsymbol{w}_0\parallel_2 = O(\sqrt{s/n})$,
	and $\sqrt{n/s} \min\limits_{1\leq j \leq s} \lvert \beta_{j0}\lvert \rightarrow \infty$.
\end{enumerate}
Under these conditions we can now state the following asymptotic distribution of the AW-DPD-LASSO with fixed weights.
\begin{theorem}\label{THM:FixW_AsymNormal}
	Suppose that, under the cases of fixed weights,  assumptions of Theorem \ref{THM:FixW_Main} hold along with Assumption (A6).  
	Then, with probability tending to one, 
	there exists a global minimizer $\widehat{\boldsymbol{\beta}} = (\widehat{\boldsymbol{\beta}}_1^o, \widehat{\boldsymbol{\beta}}_2)^T$ 
	of the AW-DPD-LASSO objective function $Q_{n,\alpha,\lambda }(\boldsymbol{\beta })$ in (\ref{1.101}), such that
	$\widehat{\boldsymbol{\beta}}_2 = \boldsymbol{0}_{p-s}$ and 
	\begin{eqnarray}
		\boldsymbol{u}^T[\boldsymbol{Z}_n^T\boldsymbol{\Omega}_n\boldsymbol{Z}_n]^{-1/2} \boldsymbol{V}_n^{-1}
		\left[\left(\widehat{\boldsymbol{\beta}}_1^o - \boldsymbol{\beta}_{10}\right) + {n\lambda_n}\boldsymbol{V}_n^2\widetilde{\boldsymbol{w}_0}\right]
		\mathop{\rightarrow}^\mathcal{L} N(0,1),
		\label{EQ:AW-DPD-LASSO_AN}
	\end{eqnarray}
	for any arbitrary $s$-dimensional vector $\boldsymbol{u}$ satisfying $\boldsymbol{u}^T\boldsymbol{u} = 1$, where  
	$\widetilde{\boldsymbol{w}_0}$ is an $s$-dimensional vector with $j$-th element $w_j sign(\beta_{j0})$.
\end{theorem}
Note that, the consistency and asymptotic distribution of the standard DPD-LASSO can now be derived from the above theorems as a special case, by defining the weight vector $\boldsymbol{w} = (1,..,1).$ It is important to note that these properties crucially depend on the structure of the oracle estimator through different assumptions on the weight vector applied to important and unimportant covariates. This dependency encourages the use of adaptive weights from the data. 

So, we now derive  the oracle and asymptotic properties of the general AW-DPD-LASSO with adaptive (stochastic)  weights based on a weight function $w(\cdot)$. We consider an initial estimator $\widetilde{\boldsymbol{\beta}}$ of the regression vector and denote $\widehat{w}_j = w(\lvert\widetilde{\beta}\lvert_j).$ 
Let the true regression parameter $\boldsymbol{\beta}_0$ produces the weight vector $\boldsymbol{w}^{\ast}=(w_1^{\ast},...,w_p^{\ast})$ and we define $${\delta}_n^\ast = \left[\sqrt{{s(\log n)}/{n}} + {\lambda_n}\left(\parallel\boldsymbol{w}_0^\ast\parallel_2 + C_2c_5\sqrt{s(\log p)/n}\right)\right],$$
where the constants $C_2, c_5$ are as defined in the following assumptions.

\begin{itemize}
	\item[(A7)]  The initial estimator $\widetilde{\boldsymbol{\beta}}$ satisfies
	$\parallel\widetilde{\boldsymbol{\beta}} - \boldsymbol{\beta}_0\parallel_2 \leq C_2 \sqrt{s(\log p)/n}$ 
	for some constant $C_2>0$, with probability tending to one.
	\item[(A8)] The weight function $w(\cdot)$ is non-increasing over $(0, \infty)$ and is Lipschitz continuous 
	with Lipschitz constant $c_5>0$. Further, 
	$w(C_2\sqrt{s(\log p)/n}) > \frac{1}{2}w(0+)$ for large enough $n$, 
	where $C_2$ is as in Assumption (A7).
	\item[(A9)] With $C_2$ being as in Assumption (A7), $\min\limits_{1\leq j \leq s}\lvert\beta_{0j}\lvert > 2C_2 \sqrt{s(\log p)/n}$.
	Further, the derivative of the wight satisfies $w'(|b|) =o(s^{-1}\lambda_n^{-1}(n\log p)^{-1/2})$ for any 
	$\lvert b \lvert > \frac{1}{2} \min\limits_{1\leq j \leq s}\lvert \beta_{0j} \lvert $.
\end{itemize}

These Assumptions (A7)-(A9) are again extremely common in the literature of adaptive penalty and hold in several practical scenarios. Particularly,  
Assumption (A7) puts a weaker restriction (only consistency)  on the initial estimator  in order to achieve the variable selection consistency of the second stage AW-DPD-LASSO estimators. 
So, it justifies the usage of  most simple LASSO estimators, including the DPD-LASSO, as the initial estimator in our proposed AW-DPD-LASSO for high-dimensional logistic regression models.  Assumptions (A8)--(A9) can also be verified for standard weight functions including the one obtained from SCAD penalty in (\ref{EQ:SCAD}).
The main results are presented in the following two theorems.

\begin{theorem}\label{THM:AdW_Main}
	Consider the cases of stochastic weights and suppose  that the assumptions of Theorem \ref{THM:FixW_Main} hold with $\boldsymbol{w}=\boldsymbol{w}^\ast$
	and $\delta_n =\delta_n^\ast$. Additionally, suppose that Assumptions (A7)-(A8) hold
	with $\lambda_n s \kappa_n \sqrt{(\log p)/n} \rightarrow 0$.
	Then, with probability tending to one, 
	there exists a global minimizer $\widehat{\boldsymbol{\beta}} = (\widehat{\boldsymbol{\beta}}_1^T, \widehat{\boldsymbol{\beta}}_2^T)^T$ 
	of the AW-DPD-LASSO objective function $Q_{n,\alpha,\lambda }(\boldsymbol{\beta})$ in (\ref{1.101}),
	with adaptive weights, such that
	$$
	\parallel\widehat{\boldsymbol{\beta}}_1 - \boldsymbol{\beta}_{10}\parallel_2 \leq C_1\delta_n,
	~~~~\mbox{and }~~~~~
	\widehat{\boldsymbol{\beta}}_2 = \boldsymbol{0}_{p-s}.
	$$ 
\end{theorem}

\begin{theorem}\label{THM:AdW_AsymNormal}
	Consider the cases of stochastic weights and suppose that the assumptions of Theorem \ref{THM:FixW_AsymNormal} hold with $\boldsymbol{w}=\boldsymbol{w}^\ast$
	and $\delta_n =\delta_n^\ast$. Additionally, suppose that Assumptions 
	\textcolor{black}{(A7)-(A9)} hold.
	Then, with probability tending to one, 
	there exists a global minimizer $\widehat{\boldsymbol{\beta}} 
	= (\widehat{\boldsymbol{\beta}}_1^T, \widehat{\boldsymbol{\beta}}_2^T)^T$ 
	of the AW-DPD-LASSO objective function $Q_{n,\alpha,\lambda }(\boldsymbol{\beta})$ in (\ref{1.101}),
	with adaptive weights, having the same asymptotic properties as those described  in Theorem \ref{THM:FixW_AsymNormal}.
\end{theorem}

\begin{remark}
	\textcolor{black}{
	It may be noted that all the theoretical results of this Section are derived under a clean data generation process as with the standard practice in the literature. This shows that the proposed AW-DPD-LASSO has nice theoretical properties as with its competitors under the  pure data cases, which is further illustrated numerically  in a particular simulation setting with pure data in Section 6 (Table 1). But, these results, more specifically the required assumptions, would not hold under the contaminated data settings; see, e.g., \cite{Chen2016}. However, that our proposed method would still provide good performances under data contamination is illustrated numerically through the contaminated simulation settings in Section 6 (Table 2-3) and real data applications in Section 7. }
	
		\textcolor{black}{Derivation of such theoretical results under contaminated settings would be an important addition to the literature, not only for our AW-DPD-LASSO but many other such existing methods, which we hope to pursue in our future research works. }
\end{remark}

\section{Computational algorithm for the proposed estimator} \label{sec5}

We present a computational algorithm using an iteratively re-weighted least squares (IRLS) approach appropriately adjusted for our DPD loss. 
\textcolor{black}{This optimization technique has been widely used, for example in \cite{pa} and \cite{Friedman}, for obtaining the penalized MLE under generalized and logistic model, respectively.}
Roughly IRLS
 uses a Newton-Raphson optimization technique where the step direction is computed by solving a weighted ordinary least squares problem by using some efficient existing algorithm. This technique has been easily extended for minimizing the penalized negative log likelihood, simply solving a penalized least squares problem at each step (\cite{Lee2}).

We first fit IRLS for the minimization of the non-penalized DPD loss function. 
More specifically, at a current solution $\boldsymbol{\beta}^{(m)}$ at $m$-th iteration, the step direction is computed as 
\begin{equation} \label{stepdirection}
	\gamma^{(m)} =  \boldsymbol{\beta}^{(m)} - \frac{\partial^2 d_\alpha(\widehat{\boldsymbol{p}},\boldsymbol{p}\left(\boldsymbol{\beta}\right))}{\partial \boldsymbol{\beta} \boldsymbol{\beta}^T} ^{-1}(\boldsymbol{\beta}^{(m)})\frac{\partial d_\alpha(\widehat{\boldsymbol{p}},\boldsymbol{p}\left(\boldsymbol{\beta}\right))}{\partial \boldsymbol{\beta}} (\boldsymbol{\beta}^{(m)}).
\end{equation}
But 
\begin{equation}\label{LambdaZ}
	\frac{\partial d_\alpha(\widehat{\boldsymbol{p}},\boldsymbol{p}\left(\boldsymbol{\beta}\right))}{\partial \boldsymbol{\beta}} (\boldsymbol{\beta}^{(m)}) = \frac{1}{n}\mathbb{X}^T\boldsymbol{H}^{(1)}_n(\boldsymbol{\beta}^{(m)}) =  \frac{1}{n} \mathbb{X}^T\boldsymbol{H}^{(2)}_n(\boldsymbol{\beta}^{(m)})\left(\boldsymbol{z}_m+\mathbb{X}\boldsymbol{\beta}^{(m)}\right)
\end{equation}
with $
\boldsymbol{z}_m = [\boldsymbol{H}^{(2)}_n(\boldsymbol{\beta}^{(m)})]^{-1}\boldsymbol{H}^{(1)}_n(\boldsymbol{\beta}^{(m)})- \mathbb{X}\boldsymbol{\beta}^{(m)},$ where matrices $\boldsymbol{H}^{(1)}_n$ and $\boldsymbol{H}^{(2)}_n$ are defined in  Section \ref{sec4}.
Thus, Equation (\ref{stepdirection}) can be rewritten as 
$$ \gamma^{(m)} = - \left( \mathbb{X}^T \boldsymbol{H}^{(2)}_n(\boldsymbol{\beta}^{(m)}) \mathbb{X}\right)^{-1} \mathbb{X}^T\boldsymbol{H}^{(2)}_n(\boldsymbol{\beta}^{(m)})\boldsymbol{z}_m. $$
From this representation, we see that the step direction can be computed by solving the least squares problem as
\begin{equation}\label{LSproblem}
	\gamma^{(m)} = - \operatorname{min}_{\boldsymbol{\gamma}} \left[(\boldsymbol{y}_m-\mathbb{X}_m\boldsymbol{\gamma})^T \cdot (\boldsymbol{y}_m-\mathbb{X}_m\boldsymbol{\gamma}) \right],
\end{equation}
with $\mathbb{X}_m = [\boldsymbol{H}^{(2)}_n(\boldsymbol{\beta}^{(m)})]^{\frac{1}{2}}\mathbb{X}$ and $\boldsymbol{y}_m = [\boldsymbol{H}^{(2)}_n(\boldsymbol{\beta}^{(m)})]^{\frac{1}{2}}\boldsymbol{z},$ i.e.,  solving a surrogate linear regression problem with design matrix $\mathbb{X}_m$ and response vector $\boldsymbol{y}_m.$
Note that the linear least squares problem does not include any intercept, but the vector $\boldsymbol{\beta}$ obtained from (\ref{stepdirection}) includes the logistic coefficient $\beta_0$ implicitly.
For penalized DPD logistic regression, it suffices to include the desirable penalization term in (\ref{LSproblem}), yielding to 
\begin{equation}\label{penLSproblem}
	\operatorname{min}_{\boldsymbol{\gamma}} \left[(\boldsymbol{y}_m-\mathbb{X}_m\boldsymbol{\gamma})^T \cdot (\boldsymbol{y}_m-\mathbb{X}_m\boldsymbol{\gamma}) + \sum_{j=1}^{k}w(\mid\tilde{\beta}_j\mid) \mid\gamma_j\mid\right]. 
\end{equation}
The penalty function for $\gamma_0$ is defined as the constant zero so the intercept term is not penalized
and the least squares estimation can be compute using efficient algorithms for adaptive LASSO. 
\cite{gm} addressed the robust estimation using adaptive LASSO procedure based in the DPD loss in ultra-high dimensional for the linear regression model, which could be also used for solving (\ref{penLSproblem}).
The previous discussion can be summarized in the following algorithm.

\noindent \textbf{Algorithm 1}[ IRLS for the computation of AW-DPD-LASSO Estimator  under the  penalized logistic regression]
\begin{enumerate}
	\item Set $m=0$. Choose values of the hyper-parameters $\lambda$ and $\alpha,$ and set a robust initial value $\boldsymbol{\beta}^{(0)}$ using any suitable robust algorithm.
	\item Set $m = m + 1 $ and do:
	\begin{itemize}
		\item Compute $\boldsymbol{H}^{(2)}_n(\boldsymbol{\beta}^{(m)})$ and $\boldsymbol{z}_m$ using (\ref{hessian}) and (\ref{LambdaZ}) respectively,
		\item Define the reweighted design matrix $\mathbb{X}_m = [\boldsymbol{H}^{(2)}_n(\boldsymbol{\beta}^{(m)})]^{1/2}\mathbb{X}$ and response vector $\boldsymbol{y}_m = [\boldsymbol{H}^{(2)}_n(\boldsymbol{\beta}^{(m)})]^{1/2}\boldsymbol{z}_m,$
		\item Solve the regularized linear regression problem (\ref{penLSproblem}) using any convenient algorithm to obtain the step direction $\boldsymbol{\gamma}^{(m)},$
		\item Update $\boldsymbol{\beta}^{(m+1)} = t \boldsymbol{\beta}^{(m)}+(1-t)\gamma^{(m)}$ by a line search over the step size $t$ to minimize the objective function (\ref{1.101}) .
	\end{itemize}
	\item If (the stopping criteria is satisfied): Stop. \\
	Else: Return to step 2.
\end{enumerate}
\begin{remark}
	\textcolor{black}{
		For each coefficient having value as zero in the preceding iteration, the associated weight is computed as 10 times the maximum of all finite weights in practical implementation of our algorithm; 
		 all weights are subsequently standardized to sum up to the number of covariates. This same process is followed for all penalties, including LASSO, which makes the resulting estimators comparable, having the same effect of the selection of weights in all cases.}
\end{remark}
Note that at each iteration of the algorithm the initial estimator of $\boldsymbol{\beta}$ is updated. Thus, our algorithm uses stochastic weights.
The previous estimation depends sharply of the value of $\lambda$. 
We apply the high-dimensional adaptation of the  Generalized Information Criterion (GIC), following the discussions in \cite{FanTang}, which is given by
\begin{equation}\label{HBIC}
	\text{HGIC}(\lambda) = \frac{ -2\log\mathcal{L}(\widehat{\boldsymbol{\beta}}_\lambda)}{n} + \frac{\log\log(n)\log(k)}{n}\parallel\widehat{\boldsymbol{\beta}}\parallel_0
\end{equation}
where $ \mathcal{L}(\widehat{\boldsymbol{\beta}}_\lambda)$ is as defined in (\ref{1.3}). 
We then select the optimal $\lambda$ by minimizing HGIC in  (\ref{HBIC}) over a predefined grid.

Our algorithm is implemented in R \footnote{\href{https://github.com/MariaJaenada/awDPDlasso}{package awDPDlasso} } , and it uses the R package \textit{glmnet} to solve the auxiliar linear regression problem (\ref{penLSproblem}).

\section{Simulation study} \label{sec6}

In this section we evaluate the performance of our proposed methods through  an extensive simulation study; \textcolor{black}{we also refer the readers to Bianco et al. (2021) for further numerical illustrations of our method and its comparisons with other competitors including the minimum penalized $\gamma$-divergence method. We have not repeated the simulation set-ups and comparisons of Bianco et al. (2021) to avoid making the present paper too lengthy. }

In the present paper, we consider an interesting simulation set-up where the response variable is generated from the logistic regression model (\ref{logit}) with $k=500$, the design matrix is drawn from a multivariate normal distribution with zero vector mean and variance-covariance matrix $\Sigma$ having Toeplitz structure, where the $(i,j)$-th element is $0.5^{|i-j|}$. \textcolor{black}{The true regression vector $\boldsymbol{\beta}_0$ is defined to be sparse, with only three non-zero components $\beta_{01} = \beta_{02} = \beta_{05}=5$  and all other components being zero (i.e., $\beta_{0j} = 0$ for all $j\neq 1,2,5$). }
In order to evaluate the robustness of our proposed methods, we additionally modify the previous set-up introducing contamination in the response variable and leverage points. For this purpose, the contaminated response data  are generated using a contaminated logistic model, 
$Y_i \sim \operatorname{Be}(\pi(\boldsymbol{x}_i^T\boldsymbol{\beta}))$ with probability $1-\varepsilon$, and $Y_i \sim \operatorname{Be}(1-\pi(\boldsymbol{x}_i^T\boldsymbol{\beta}))$ with probability $\varepsilon$, where $\varepsilon$ represent the contamination level. 
In case of leverage points, we introduce contamination in observations with response $Y=1$ by adding to one variable with true non-zero coefficient (randomly chosen for each sample) a random variable generated from  a normal distribution, $\mathcal{N}\left(-5,0.01\right)$ or to five variables with true zero coefficient (randomly chosen for each sample) a random variable generated from $\mathcal{N}\left(5,0.01\right)$. Contaminated observations of both types are selected randomly up to a total of $\varepsilon = 5, 10\%$ of the samples

The performance of our method is studied and compared via  different error measures. Particularly, model size (MS), true positive proportion (TP) and true negative proportion (TN) are used to evaluate the variable selection accuracy, whereas  the mean square error for the true non-zero coefficients (MSES) and the mean absolute error (MAE) of the estimated parameters are used to evaluate the accuracy on the estimation.
Formally, these measures are defined as 
$$
\operatorname{MS}(\widehat{\boldsymbol{\beta}}) = \text{supp}(\widehat{\boldsymbol{\beta}}),
\operatorname{TP}(\widehat{\boldsymbol{\beta}}) = \frac{\text{supp}(\widehat{\boldsymbol{\beta}})\cap\text{supp}(\boldsymbol{\beta}_0)}{\text{supp}(\boldsymbol{\beta}_0)},  
\operatorname{TN}(\widehat{\boldsymbol{\beta}}) = \frac{\text{supp}^c(\widehat{\boldsymbol{\beta}})\cap\text{supp}^c(\boldsymbol{\beta}_0)}{\text{supp}^c(\boldsymbol{\beta}_0)}.
$$
$$
\operatorname{MSES}(\widehat{\boldsymbol{\beta}}) = \frac{1}{s} \parallel \widehat{\boldsymbol{\beta}}_{\mathcal{S}} - \boldsymbol{\beta}_{0\mathcal{S}} \parallel^2,  \hspace{0.3cm}
\operatorname{MAE}(\widehat{\boldsymbol{\beta}}) = \frac{1}{p} \parallel \widehat{\boldsymbol{\beta}} - \boldsymbol{\beta}_{0} \parallel_1.
$$

For comparison, we also fit the logistic model with the classical \textcolor{black}{LASSO penalized MLE}  (LASSO) \textcolor{black}{and its adaptive versions penalized with the hard-thresholding (\textcolor{black}{Ad-LASSO}) and the SCAD penalty (AW-LASSO).}
We also considered some other existing robust methods, namely \textcolor{black}{the robust elastic net- penalized estimator (LTS) of (\cite{parkkonishi}) based on Mahalanobis distance} and \textcolor{black}{the robust M-estimator corresponding to the Hubarized residual based  quasilikelihood loss function (RLASSO) proposed  by \cite{AvellaRonchetti} along with its adaptive version (Ad-RLASSO)}. Further, we consider  three different weight functions in our proposal, corresponding to the DPD-LASSO, Ad-DPD-LASSO and a general formulation of AW-DPD-LASSO with weight function as in (\ref{EQ:SCAD}) in Remark \ref{remarkscad} of Section \ref{sec2}. 

To compute LASSO, Ad-LASSO and AW-LASSO we use the R package \textit{glmnet}. The penalty tuning parameter $\lambda$ is chosen by cross-validation using the implemented function \textit{cv.glmenet}. As the AW-LASSO penalty depends on $\lambda$, the penalty weights are computed over a grid of $\lambda,$ and then we have applied cross-validation for choosing the best penalty parameter for each weight vector.
The LTS method is carried out using the R package  \textit{enetLTS}, whereas RLASSO and Ad-RLASSO are computed as per the  implementation proposed by the authors but using HGIC criterion to select the optimal $\lambda$. Our algorithm is also implemented in R, and uses the R package \textit{glmnet} to solve the auxiliar linear regression problem (\ref{penLSproblem}). \textcolor{black}{We would like to point out that, because the computational estimation algorithms are differently implemented, there may be small variations on the results due to implementation as is the case with most high-dimensional situations. Nevertheless, the packages \textit{glmnet} and \textit{enetLTS}  are well established and used throughout the scientific community for penalized regression estimation and we believe that improving the algorithm is hardly achievable.}


Tables \ref{resulstclean}-\ref{results10} show the performance evaluation measures produced by the different methods considered in our simulation exercises. All estimators perform well and competitively  in the absence of contamination, and it is straightforward to see that adaptive methods enhance the variable selection property. Further, our proposed algorithm decreases the estimation error and rejects a major number of insignificant variables, specially with weighted penalties. In a contaminated scenario, the results show a clear significant improvement on the estimation accuracy and robustness. Non-robust adaptive method Ad-LASSO dismiss significant variables, while Ad-DPD-LASSO and AW-DPD-LASSO maintain the rate of TP. Moreover, the improvement in the model selection and estimation accuracy is preserved in all settings. The LTS method selects the largest number of variables, but has the best TP rate in the presence of contaminated data. On the other hand, RLASSO and Ad-RLASSO estimators lack robustness under leverage points.

Besides, there exists a considerable difference between Ad-DPD-LASSO and AW-DPD-LASSO estimators. The AW-DPD-LASSO generally selects a greater number of variables, containing mostly the truly important variables, but produces lower error in estimation.

\textcolor{black}{
	Further, it may be noted there there is some effect of the choice of the robustness tuning parameter $\alpha$ for our proposed procedure with any given weights. For example, as seen particularly in Table \ref{results5}, TP of AW-DPD-LASSO decrease as $\alpha$ increases and this effect is slightly greater for the AW-DPD-LASSO compared to the DPD-LASSO with respect to the variable selection. However, these changes in the performances of the proposed estimators in terms of varying $\alpha$ is comparatively less pronounced  than the associated effects under classical low-dimensional model set-ups. So, rather than  choosing a data driven $\alpha$ in the present high-dimensional context, which would consist of an involved parameter selection strategy, one quick and simple solution would be to use any $\alpha$ value in the range  [0.3, 0.5] to get reasonably good performance of the proposed procedure. This fact and suggestion are indeed consistent with previous works with DPD-based loss functions under high-dimensional settings (e.g., \cite{ghosh20, gm}).
	However, if one does wish to use an optimal  data-driven choice of $\alpha>0$, the standard existing procedures under the low-dimensional context (e.g., \cite{warwick2005}; \cite{ghoshbasu2016}, \cite{basak2021}, etc.) may be extended to the high-dimensional context at some additional  computational cost. Note that, in the present context, there is another tuning parameter, namely $\lambda$, which is also plying a crucial role in the variable selection results for each $\alpha>0$.  In fact, the choice of $\alpha$ and $\lambda$ together gives the finite-sample results and the individual effects are not easy identify: more in-depth investigations would be required to study the individual effects of both the tuning parameters $\alpha$ and $\lambda$. This (together with further study of the parameter estimates) we hope to pursue in our future research.
}

\begin{table}
	\caption{\label{resulstclean} Error measures obtained by  different methods under pure data.}
	\centering
	\fbox{%
		\begin{tabular}{lccccc}
			& MS & TP & TN & MSES & MAE \\ 
			& &  &  &  & $\times 10^2$ \\
			\hline
			LASSO & 15.41 & 1.00 & 0.98 & 16.07 & 2.63 \\ 
			Ad-LASSO & 5.56 & 1.00 & 0.99 & 6.25 & 1.54 \\
			AW-LASSO & 3.03 & 0.71 & 1.00 & 13.41 & 1.99  \\   
			LTS & 29.34 & 1.00 & 0.95 & 11.68 & 2.78 \\ 
			RLASSO & 4.32 & 1.00 & 1.00 & 10.37 & 1.94 \\ 
			Ad-RLASSO & 4.18 & 1.00 & 1.00 & 3.83 & 1.17 \\ 
			\hline
			DPD-LASSO $\alpha = $ 0.1 & 14.69 & 1.00 & 0.98 & 15.70 & 2.59 \\ 
			DPD-LASSO $\alpha = $ 0.3 & 15.28 & 1.00 & 0.98 & 16.03 & 2.63 \\ 
			DPD-LASSO $\alpha = $ 0.5 & 15.40 & 1.00 & 0.98 & 16.07 & 2.63 \\ 
			DPD-LASSO $\alpha = $ 0.7 & 15.41 & 1.00 & 0.98 & 16.07 & 2.63 \\ 
			DPD-LASSO $\alpha = $ 1 & 15.41 & 1.00 & 0.98 & 16.28 & 2.63 \\ 
			\hline
			Ad-DPD-LASSO $\alpha = $0.1 & 4.30 & 1.00 & 1.00 & 3.42 & 1.15 \\ 
			Ad-DPD-LASSO $\alpha = $0.3 & 4.51 & 1.00 & 1.00 & 3.59 & 1.21 \\ 
			Ad-DPD-LASSO $\alpha = $0.5 & 4.57 & 1.00 & 1.00 & 4.00 & 1.27 \\ 
			Ad-DPD-LASSO $\alpha = $0.7 & 4.65 & 1.00 & 1.00 & 4.43 & 1.35 \\ 
			Ad-DPD-LASSO $\alpha = $1 & 4.66 & 1.00 & 1.00 & 4.90 & 1.41 \\ 
			\hline
			AW-DPD-LASSO $\alpha = $0.1 & 4.17 & 1.00 & 1.00 & 1.20 & 0.69 \\ 
			AW-DPD-LASSO $\alpha = $0.3 & 4.25 & 1.00 & 1.00 & 1.07 & 0.68 \\ 
			AW-DPD-LASSO $\alpha = $0.5 & 4.25 & 1.00 & 1.00 & 1.13 & 0.69 \\ 
			AW-DPD-LASSO $\alpha = $0.7 & 4.23 & 1.00 & 1.00 & 1.26 & 0.71 \\ 
			AW-DPD-LASSO $\alpha = $1 & 4.21 & 1.00 & 1.00 & 1.34 & 0.72 \\ 
	\end{tabular}}
\end{table}

\begin{table}
	\caption{\label{results5} Error measures obtained by  different methods under  $5\%$ contaminated data.}
	\centering
	\fbox{%
		\begin{tabular}{lccccc}
			& Size & TP & TN & MSES$(\widehat{\boldsymbol{\beta}})$  & MAE$(\widehat{\boldsymbol{\beta}})$  \\ 
			& &  &  & $\times 10^2$ & \\
			\hline
			\multicolumn{6}{c}{$5\%$ of outliers at the response variable} \\
			\hline
			LASSO & 9.77 & 0.98 & 0.99 & 20.18 & 2.81 \\ 
			Ad-LASSO & 5.26 & 0.96 & 1.00 & 15.55 & 2.47 \\ 
			AW-LASSO & 4.32 & 0.73 & 1.00 & 17.88 & 2.56 \\ 
			LTS & 29.76 & 0.99 & 0.95 & 13.98 & 3.10 \\ 
			RLASSO & 3.80 & 0.85 & 1.00 & 15.59 & 2.38 \\ 
			AdRLASSO & 3.75 & 0.85 & 1.00 & 10.32 & 1.89 \\ 
			\hline
			DPD-LASSO $\alpha = $0.1 & 9.52 & 0.98 & 0.99 & 19.84 & 2.79 \\ 
			DPD-LASSO $\alpha = $0.3 & 9.64 & 0.98 & 0.99 & 20.03 & 2.80 \\ 
			DPD-LASSO $\alpha = $0.5 & 9.69 & 0.98 & 0.99 & 20.10 & 2.80 \\ 
			DPD-LASSO $\alpha = $0.7 & 9.70 & 0.98 & 0.99 & 20.13 & 2.81 \\ 
			DPD-LASSO $\alpha = $1 & 9.71 & 0.98 & 0.99 & 20.15 & 2.81 \\ 
			\hline
			Ad-DPD-LASSO $\alpha = $0.1 & 4.60 & 0.97 & 1.00 & 10.93 & 2.12 \\ 
			Ad-DPD-LASSO $\alpha = $0.3 & 4.68 & 0.97 & 1.00 & 10.80 & 2.11 \\ 
			Ad-DPD-LASSO $\alpha = $0.5 & 4.48 & 0.97 & 1.00 & 11.00 & 2.10 \\ 
			Ad-DPD-LASSO $\alpha = $0.7 & 4.78 & 0.97 & 1.00 & 10.55 & 2.12 \\ 
			Ad-DPD-LASSO $\alpha = $1 & 4.64 & 0.97 & 1.00 & 10.89 & 2.11 \\ 
			\hline
			AW-DPD-LASSO $\alpha = $0.1 & 5.01 & 0.97 & 1.00 & 7.19 & 1.98 \\ 
			AW-DPD-LASSO $\alpha = $0.3 & 5.05 & 0.97 & 1.00 & 5.64 & 1.82 \\ 
			AW-DPD-LASSO $\alpha = $0.5 & 5.03 & 0.95 & 1.00 & 5.67 & 1.79 \\ 
			AW-DPD-LASSO $\alpha = $0.7 & 4.84 & 0.94 & 1.00 & 6.37 & 1.77 \\ 
			AW-DPD-LASSO $\alpha = $1 & 4.90 & 0.94 & 1.00 & 6.55 & 1.78 \\ 
			\hline
			\multicolumn{6}{c}{ $5\%$ of leverage points } \\
			\hline
			LASSO & 8.59 & 0.91 & 0.99 & 20.90 & 2.84 \\ 
			Ad-LASSO & 4.90 & 0.85 & 1.00 & 16.93 & 2.56 \\ 
			AW-LASSO & 3.40 & 0.65 & 1.00 & 19.20 &  2.50  \\ 
			LTS & 28.00 & 0.93 & 0.95 & 14.71 & 2.97 \\  
			RLASSO & 3.25 & 0.71 & 1.00 & 16.67 & 2.43 \\  
			Ad-RLASSO & 3.23 & 0.71 & 1.00 & 13.56 & 2.16 \\ 
			\hline
			DPD-LASSO $\alpha = $0.1 & 8.22 & 0.92 & 0.99 & 20.61 & 2.81 \\ 
			DPD-LASSO $\alpha = $0.3 & 8.43 & 0.92 & 0.99 & 20.74 & 2.82 \\ 
			DPD-LASSO $\alpha = $0.5 & 8.48 & 0.91 & 0.99 & 20.81 & 2.83 \\ 
			DPD-LASSO $\alpha = $0.7 & 8.51 & 0.91 & 0.99 & 20.81 & 2.83 \\ 
			DPD-LASSO $\alpha = $1 & 8.50 & 0.91 & 0.99 & 20.81 & 2.83 \\ 
			\hline
			Ad-DPD-LASSO $\alpha = $0.1 & 4.52 & 0.90 & 1.00 & 12.68 & 2.27 \\ 
			Ad-DPD-LASSO $\alpha = $0.3 & 4.43 & 0.89 & 1.00 & 13.02 & 2.28 \\ 
			Ad-DPD-LASSO $\alpha = $0.5 & 4.24 & 0.88 & 1.00 & 13.41 & 2.28 \\ 
			Ad-DPD-LASSO $\alpha = $0.7 & 4.25 & 0.87 & 1.00 & 13.66 & 2.31 \\ 
			Ad-DPD-LASSO $\alpha = $1 & 4.16 & 0.87 & 1.00 & 14.05 & 2.33 \\ 
			\hline
			AW-DPD-LASSO $\alpha = $0.1 & 5.46 & 0.93 & 0.99 & 8.33 & 2.23 \\ 
			AW-DPD-LASSO $\alpha = $0.3 & 5.86 & 0.89 & 0.99 & 8.42 & 2.49 \\ 
			AW-DPD-LASSO $\alpha = $0.5 & 5.07 & 0.84 & 0.99 & 9.85 & 2.33 \\ 
			AW-DPD-LASSO $\alpha = $0.7 & 4.83 & 0.83 & 1.00 & 10.57 & 2.30 \\ 
			AW-DPD-LASSO $\alpha = $1 & 4.82 & 0.82 & 1.00 & 11.11 & 2.29 \\ 
		\end{tabular}
	}
\end{table}

\begin{table}
	\caption{\label{results10}Error measures obtained by  different methods under  $10\%$ contaminated data.}
	\centering
	\fbox{%
		\begin{tabular}{lccccc}
			& Size & TP & TN & MSES$(\widehat{\boldsymbol{\beta}})$  & MAE$(\widehat{\boldsymbol{\beta}})$  \\ 
			& &  &  & $\times 10^2$ & \\
			\hline
			\multicolumn{6}{c}{$10\%$ of outliers at the response variable} \\
			\hline
			& Size & TP & TN & MSES & MAE \\ 
			\hline
			LASSO & 7.93 & 0.87 & 0.99 & 22.06 & 2.91 \\ 
			Ad-LASSO & 5.08 & 0.82 & 0.99 & 18.96 & 2.75 \\ 
			AW-LASSO & 3.83& 0.45 & 0.99 & 22.52 & 2.90 \\ 
			LTS & 27.45 & 0.93 & 0.95 & 16.89 & 3.25 \\ 
			RLASSO & 3.06 & 0.63 & 1.00 & 19.43 & 2.66 \\ 
			Ad-RLASSO & 3.05 & 0.63 & 1.00 & 16.37 & 2.43 \\ 
			\hline
			DPD-LASSO $\alpha = $0.1 & 7.79 & 0.87 & 0.99 & 21.80 & 2.90 \\ 
			DPD-LASSO $\alpha = $0.3 & 7.83 & 0.87 & 0.99 & 21.89 & 2.90 \\ 
			DPD-LASSO $\alpha = $0.5 & 7.86 & 0.87 & 0.99 & 21.96 & 2.90 \\ 
			DPD-LASSO $\alpha = $0.7 & 7.86 & 0.87 & 0.99 & 21.98 & 2.91 \\ 
			DPD-LASSO $\alpha = $1 & 7.86 & 0.87 & 0.99 & 22.00 & 2.91 \\ 
			\hline
			Ad-DPD-LASSO $\alpha = $0.1 & 4.77 & 0.86 & 1.00 & 15.20 & 2.57 \\ 
			Ad-DPD-LASSO $\alpha = $0.3 & 4.45 & 0.85 & 1.00 & 15.77 & 2.54 \\ 
			Ad-DPD-LASSO $\alpha = $0.5 & 4.42 & 0.84 & 1.00 & 15.72 & 2.53 \\ 
			Ad-DPD-LASSO $\alpha = $0.7 & 4.27 & 0.84 & 1.00 & 16.03 & 2.53 \\ 
			Ad-DPD-LASSO $\alpha = $1 & 4.23 & 0.84 & 1.00 & 15.89 & 2.51 \\ 
			\hline
			AW-DPD-LASSO $\alpha = $0.1 & 6.16 & 0.91 & 0.99 & 10.12 & 2.74 \\ 
			AW-DPD-LASSO $\alpha = $0.3 & 6.06 & 0.90 & 0.99 & 9.92 & 2.82 \\ 
			AW-DPD-LASSO $\alpha = $0.5 & 5.07 & 0.85 & 0.99 & 11.43 & 2.49 \\ 
			AW-DPD-LASSO $\alpha = $0.7 & 4.77 & 0.83 & 1.00 & 12.40 & 2.45 \\ 
			AW-DPD-LASSO $\alpha = $1 & 4.45 & 0.82 & 1.00 & 13.20 & 2.39 \\ 
			\hline
			\multicolumn{6}{c}{ $10\%$ of leverage points } \\
			\hline
			LASSO & 7.53 & 0.72 & 0.99 & 21.62 & 2.89 \\ 
			Ad-LASSO & 4.90 & 0.69 & 0.99 & 18.60 & 2.72 \\ 
			AW-LASSO & 3.58 & 0.74 & 1.00 & 17.37 & 2.50 \\ 
			LTS & 26.58 & 0.77 & 0.95 & 16.17 & 3.09 \\  
			RLASSO & 3.09 & 0.63 & 1.00 & 16.93 & 2.45 \\ 
			Ad-RLASSO & 3.09 & 0.63 & 1.00 & 16.40 & 2.44 \\ 
			\hline
			DPD-LASSO $\alpha = $0.1 & 7.26 & 0.72 & 0.99 & 21.29 & 2.86 \\ 
			DPD-LASSO $\alpha = $0.3 & 7.41 & 0.72 & 0.99 & 21.44 & 2.87 \\ 
			DPD-LASSO $\alpha = $0.5 & 7.46 & 0.72 & 0.99 & 21.51 & 2.88 \\ 
			DPD-LASSO $\alpha = $0.7 & 7.45 & 0.72 & 0.99 & 21.54 & 2.88 \\ 
			DPD-LASSO $\alpha = $1 & 7.47 & 0.72 & 0.99 & 21.56 & 2.88 \\ 
			\hline
			Ad-DPD-LASSO $\alpha = $0.1 & 3.84 & 0.72 & 1.00 & 15.80 & 2.49 \\ 
			Ad-DPD-LASSO $\alpha = $0.3 & 3.89 & 0.71 & 1.00 & 16.14 & 2.53 \\ 
			Ad-DPD-LASSO $\alpha = $0.5 & 3.73 & 0.70 & 1.00 & 16.52 & 2.53 \\ 
			Ad-DPD-LASSO $\alpha = $0.7 & 3.77 & 0.70 & 1.00 & 16.57 & 2.54 \\ 
			Ad-DPD-LASSO $\alpha = $1 & 3.74 & 0.70 & 1.00 & 16.70 & 2.54 \\ 
			\hline
			AW-DPD-LASSO $\alpha = $0.1 & 5.40 & 0.76 & 0.99 & 11.59 & 2.70 \\ 
			AW-DPD-LASSO $\alpha = $0.3 & 5.51 & 0.76 & 0.99 & 11.55 & 2.81 \\ 
			AW-DPD-LASSO $\alpha = $0.5 & 4.74 & 0.73 & 0.99 & 12.73 & 2.68 \\ 
			AW-DPD-LASSO $\alpha = $0.7 & 4.37 & 0.72 & 1.00 & 13.51 & 2.59 \\ 
			AW-DPD-LASSO $\alpha = $1 & 3.90 & 0.71 & 1.00 & 14.13 & 2.44 \\
		\end{tabular}
	}
\end{table}

\section{Real Data Applications: Genomic Classification of Cancer Patients } \label{sec7}

\subsection{Breast Cancer}

We apply our methods to two different datasets regarding breast cancer. The first dataset is from  \cite{Veer}, which contains $p=24481$ gene expression levels of $n=78$ breast cancer patients, 34 of whom had developed distance metastasis within 5 years and 44 continued to be disease-free after that period. All cancer samples have the same histology, and from each patient, $5\mu g$ total RNA was isolated from snap-frozen tumour material and used to derive complementary RNA. 
A reference cRNA pool was made by pooling equal amounts of cRNA from each of the sporadic carcinomas. Two hybridizations were carried out for each tumour using a fluorescent dye reversal technique on microarrays containing approximately 25.000 human genes synthesized. Fluorescence intensities of scanned images were quantified, normalized, and corrected to yield the transcript abundance of a gene as an intensity ratio with respect to that of the signal of the reference pool. Some 5.000 genes were significantly regulated across the group of samples. Moreover, 19 additional breast cancer patients'  gene expressions are used as test data.
The aim is to discriminate between relapse and non-relapse observations.

We first delete genes with null variation through samples, and apply to the resultant microarray data a feature selection by ranking the genes by their correlation to the response variable, and selecting the genes which have correlations higher than 0.3 with the class distinction. This feature selection reduces the number of explanatory genes to $p=457.$ We also scale the data so all gene expressions have zero mean and unit variance. According to \cite{Zervakis}, observations numbered  37, 38, 54, 60 and 76 can be characterized as outliers and hence, we should study the influence of these five outlying observations on the performance of the model on the rest of the sample.

\begin{figure}[htb]
	\centering
		\begin{tabular}{cc}
		\begin{subfigure}{0.5\textwidth}
				\includegraphics[scale=0.35]{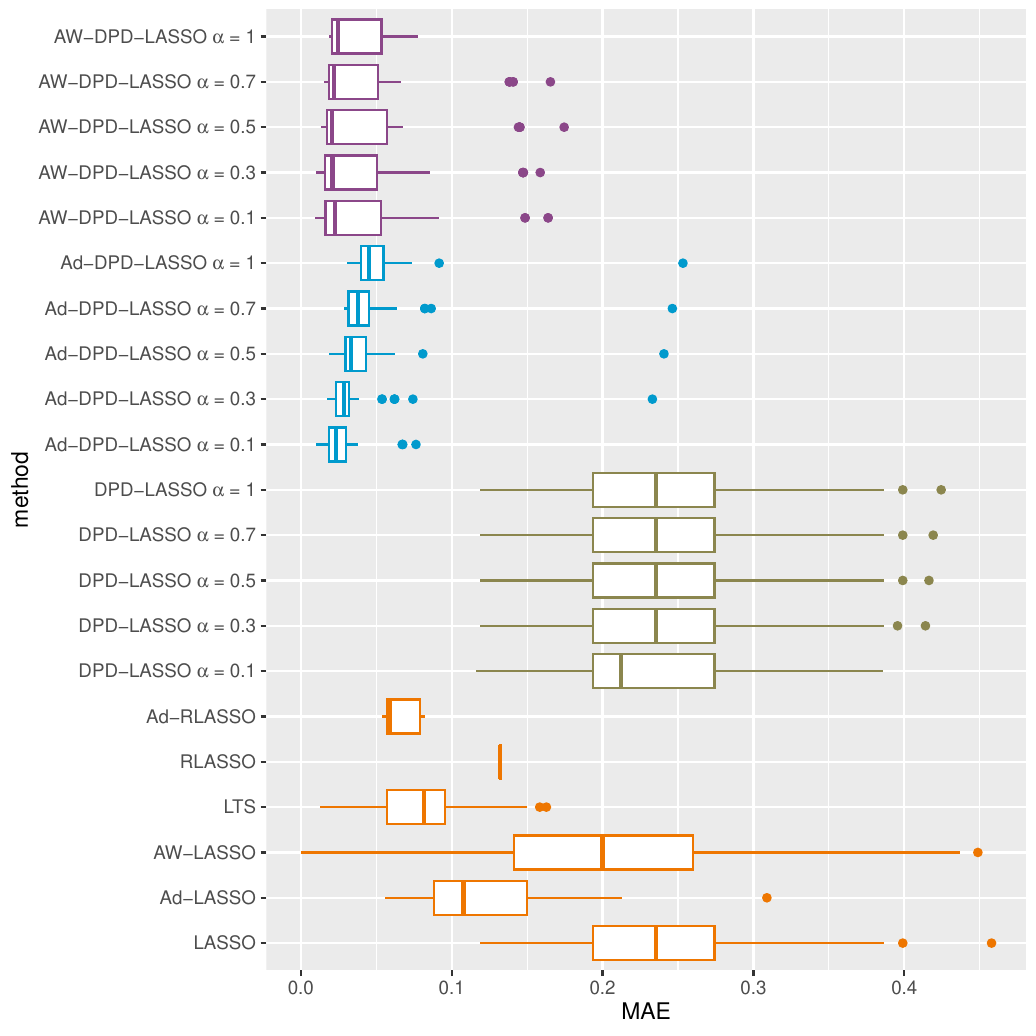}
				\subcaption{\textcolor{black}{Breast Cancer \cite{Veer}} }
				\label{fig:boxplotBC}
			\end{subfigure}
		&
		\begin{subfigure}{0.5\textwidth}
				\includegraphics[scale=0.35]{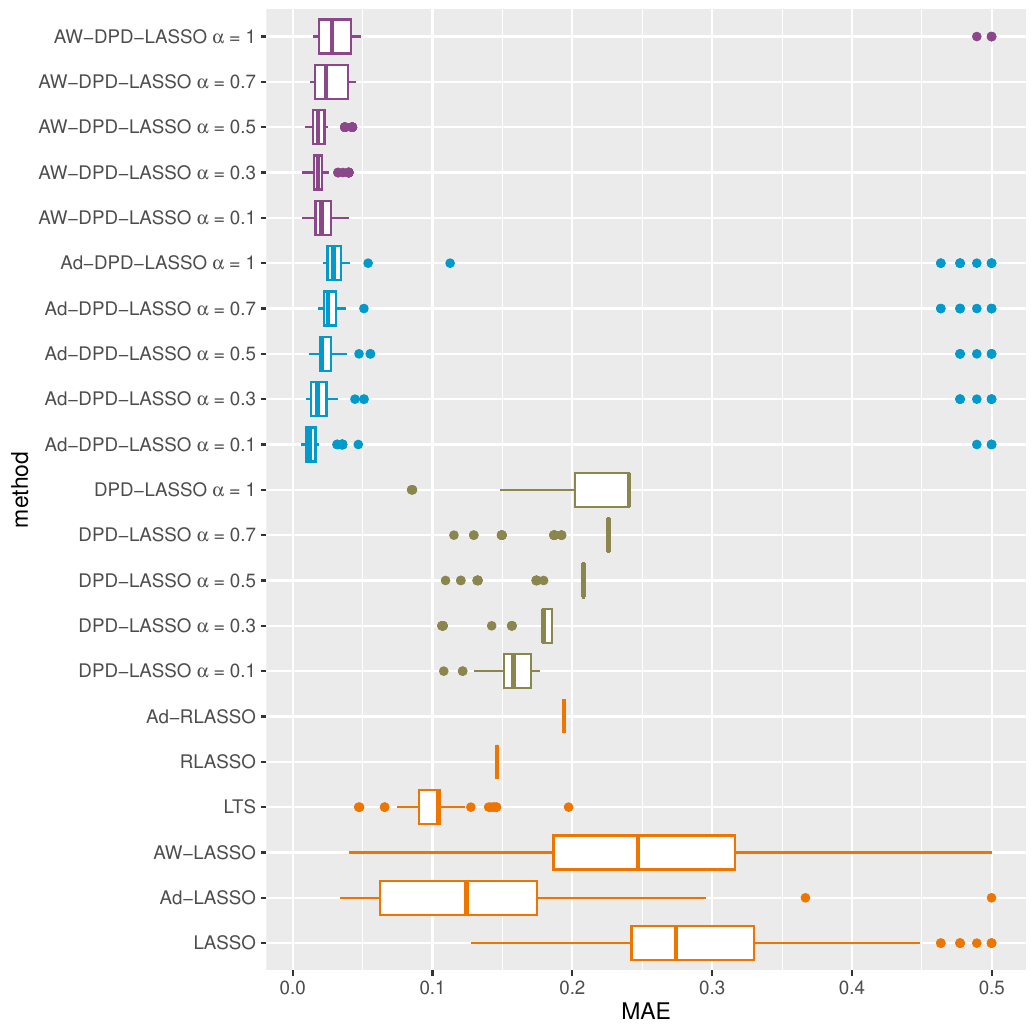}
				\subcaption{\textcolor{black}{Breast Cancer \cite{West}}}
				\label{fig:boxplotBC2}
			\end{subfigure}\\
		\begin{subfigure}{0.5\textwidth}
				\includegraphics[scale=0.35]{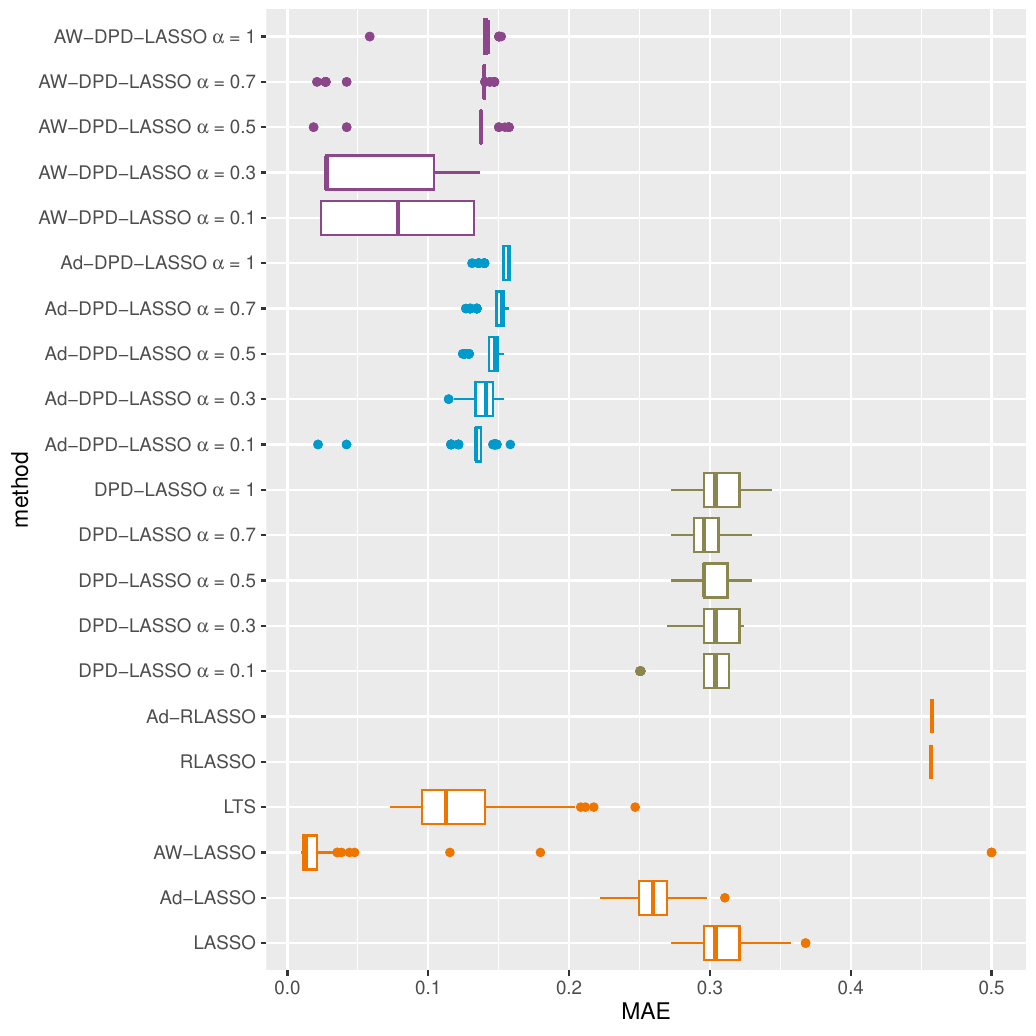} 
				\subcaption{\textcolor{black}{Colon tumour \cite{Alon1999}}}
				\label{fig:boxplotC}
			\end{subfigure}
		&
		\begin{subfigure}{0.5\textwidth}
				\includegraphics[scale=0.35]{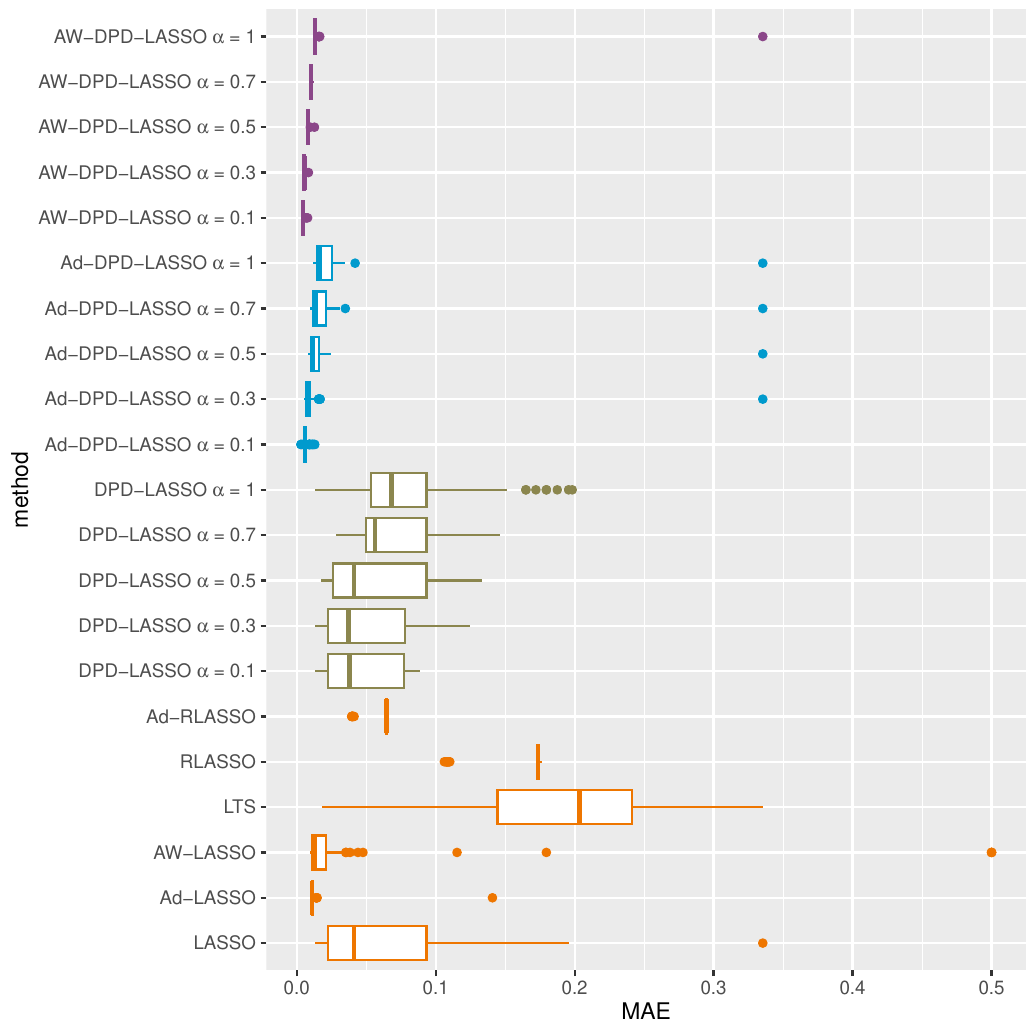}
				\subcaption{\textcolor{black}{Leukemia \cite{golub}}}
				\label{fig:boxplotL}
			\end{subfigure}
		\end{tabular}
		\caption{Boxplot of the mean square error (MAE) produced with the different methods in $R=100$ fits of the real datasets.}
		\label{boxplot}
	\end{figure}

\textcolor{black}{
	On the other hand, as discussed in Section \ref{sec6}, the estimation of the regression coefficients is  carried out by applying numerical optimization methods  to all practical datasets.  These numerical methods rely on a stochastic component, which causes the estimated logistic models to differ  slightly (both in variable selection and parameter estimation) every time the algorithm is run on the same dataset. So, to fairly compare the accuracy of the estimation methods  in our real data applications, we reports all the error measures averaged over the logistic model fit obtained in  $R=100$ different runs for each real datasets. In particular, for the present breast cancer dataset, 
}
Table \ref{tablebreast1} shows the mean model size (MS) and the mean accuracy of the different methods with the unclean data (training set) as well as the cleaned data after removal of outliers and also the test data. The accuracy of the methods with cleaned and test data is measured using the fitted model with the original dataset.
DPD-based methods are initialized with the LASSO estimator as in the simulation study. As illustrated, the number of selected genes is lower with the adaptive methods, around 11-14 genes in contrast to the standard LASSO RLASSO and LTS methods, which select over 20 and 37 genes, respectively. Conversely, the highest accuracy with cleaned data is achieved with Ad-RLASSO jointly with the proposed robust methods, Ad-DPD-LASSO and AW-DPD-LASSO and in test data with RLASSO, LTS and the proposed Ad-DPD-LASSO methods, indicating a slight over-fitting of the AW-DPD-LASSO. 
Figure \ref{fig:boxplotBC} (top left)  shows a boxplot of the mean absolute error (MAE) between the predicted probabilities and real classes for all methods based on $R=100$ iterations. The superiority of the Ad-DPD-LASSO and AW-DPD-LASSO is obvious, which indicates better predictive performance of these methods for the present dataset

\begin{table}
	\caption{\label{tablebreast1} Model size and accuracy of the different methods with unclean data (training set), cleaned data (after outlier removal) and test data for \cite{Veer}  breast cancer dataset.}
	\centering
	\fbox{
		\begin{tabular}{lrrrr}
			& & \multicolumn{3}{c}{Accuracy ($\%$)} \\
			\cline{3-5}
			& MS &  unclean &  cleaned  & test \\ 
			\hline
			LASSO & 26.09 & 96.83 & 98.12 & 63.53 \\ 
			Ad-LASSO & 13.35 & 96.69 & 97.49 & 65.21 \\ 
			AW-LASSO & 15.73 & 92.23 & 93.56 & 62.68 \\
			RLASSO  &  20.00 & & 69.00 &\\
			AD-RLASSO & 12.73 &  & 73.00 & \\
			LTS & 37.38 & 93.09 & 93.75 & 74.74 \\ 
			DPD-LASSO  $\alpha = $ 0.1 & 25.51 & 98.01 & 98.70 & 68.16 \\ 
			DPD-LASSO  $\alpha = $ 0.3 & 26.10 & 97.56 & 98.59 & 65.47 \\ 
			DPD-LASSO  $\alpha = $ 0.5 & 26.19 & 97.04 & 98.32 & 63.74 \\ 
			DPD-LASSO  $\alpha = $ 0.7 & 26.18 & 97.00 & 98.28 & 63.63 \\ 
			DPD-LASSO  $\alpha = $ 1 & 26.17 & 96.99 & 98.26 & 63.74 \\ 
			Ad-DPD-LASSO  $\alpha = $ 0.1 & 14.06 & 99.90 & 99.88 & 72.95 \\ 
			Ad-DPD-LASSO  $\alpha = $ 0.3 & 12.07 & 99.73 & 99.74 & 73.47 \\ 
			Ad-DPD-LASSO  $\alpha = $ 0.5 & 11.90 & 99.73 & 99.74 & 74.74 \\ 
			Ad-DPD-LASSO  $\alpha = $ 0.7 & 11.76 & 99.74 & 99.75 & 72.89 \\ 
			Ad-DPD-LASSO  $\alpha = $ 1 & 11.16 & 99.73 & 99.74 & 74.63 \\ 
			AW-DPD-LASSO  $\alpha = $ 0.1 & 11.57 & 98.46 & 98.65 & 64.58 \\ 
			AW-DPD-LASSO  $\alpha = $ 0.3 & 11.46 & 98.60 & 98.68 & 64.89 \\ 
			AW-DPD-LASSO  $\alpha = $ 0.5 & 11.30 & 98.64 & 98.67 & 66.05 \\ 
			AW-DPD-LASSO  $\alpha = $ 0.7 & 11.27 & 98.83 & 98.90 & 66.32 \\ 
			AW-DPD-LASSO  $\alpha = $ 1 & 11.39 & 99.33 & 99.39 & 66.68 \\ 
	\end{tabular}}
\end{table}

\begin{table}
	\caption{\label{tablebreast2} Model size and accuracy of the different methods with unclean data (training set) and cleaned data (after outlier removal) for \cite{West}  breast cancer dataset .}
	\centering
	\fbox{
		\begin{tabular}{lrrr}
			& & \multicolumn{2}{c}{Accuracy ($\%$)} \\
			\cline{3-4}
			& MS &  unclean &  cleaned  \\ 
			\hline
			LASSO & 14.74 & 97.73 & 97.62 \\ 
			Ad-LASSO & 6.74 & 98.04 & 97.72 \\ 
			AW-LASSO & 14.35 & 98.20 & 97.92 \\
			LTS & 24.60 & 90.86 & 91.25 \\ 
			RLASSO & 19.00 & 100.00 & 100.00 \\ 
			AD-RLASSO & 9.00 & 100.00 & 100.00 \\ 
			DPD-LASSO $\alpha = $ 0.1 & 12.18 & 100.00 & 100.00 \\ 
			DPD-LASSO $\alpha = $ 0.3 & 12.39 & 100.00 & 100.00 \\ 
			DPD-LASSO $\alpha = $ 0.5 & 12.59 & 100.00 & 100.00 \\ 
			DPD-LASSO $\alpha = $ 0.7 & 12.70 & 100.00 & 100.00 \\ 
			DPD-LASSO $\alpha = $ 1 & 13.11 & 100.00 & 100.00 \\ 
			Ad-DPD-LASSO $\alpha = $ 0.1 & 7.95 & 98.31 & 98.33 \\ 
			Ad-DPD-LASSO $\alpha = $ 0.3 & 7.20 & 98.08 & 98.05 \\ 
			Ad-DPD-LASSO $\alpha = $ 0.5 & 6.59 & 98.10 & 98.08 \\ 
			Ad-DPD-LASSO $\alpha = $ 0.7 & 6.28 & 98.02 & 97.97 \\ 
			Ad-DPD-LASSO $\alpha = $ 1 & 6.15 & 97.98 & 97.92 \\ 
			AW-DPD-LASSO $\alpha = $ 0.1 & 7.26 & 100.00 & 100.00 \\ 
			AW-DPD-LASSO $\alpha = $ 0.3 & 7.25 & 100.00 & 100.00 \\ 
			AW-DPD-LASSO $\alpha = $ 0.5 & 6.99 & 100.00 & 100.00 \\ 
			AW-DPD-LASSO $\alpha = $ 0.7 & 6.97 & 99.96 & 99.95 \\ 
			AW-DPD-LASSO $\alpha = $ 1 & 6.72 & 98.29 & 98.30 \\ 
	\end{tabular}}
\end{table}

Additionally, we consider the breast cancer dataset investigated in  \cite{West}.
These data contain gene expression profiles of 49 patients with breast cancer, 25 estrogen receptor (ER+) tumours (cells of this type of breast cancer have receptors that allow them to use the hormone estrogen to grow) and 24  non estrogen receptors breast cancers (ER-). All assays used the human HuGeneFL GENECHIP microarray. Arrays were hybridized and DNA chips were scanned with the GENECHIP scanner and processed by GENECHIP Expression Analysis algorithm. All patients had the same histology (invasive ductal carcinoma) and each tumour was between 1.5 and 5 cm in maximal dimension. The aim is to distinguish between ER+ and ER- tumours.

These data have been widely studied in the cancer classification field, specially for mislabelling detection algorithms. \cite{West} pointed out there is biological evidence that observations 14, 16, 31, 33, 40, 43, 45 and 46 are mislabelled, and observation 11 is clearly characterized as unusual. The original data consisted of $p=7129$ gene expression levels, which were reduced to the $p= 799$ most correlated genes with the class distinction as in the previous dataset. We scale the data so that all gene expression levels have zero mean and unit variance. We could asses the robustness of the estimators evaluating the estimated models fitted with the original contaminated dataset in the cleaned data (without outliers) and comparing the performance with the different estimating methods. Table \ref{tablebreast2} contains the mean model size and mean accuracy obtained with the different estimating methods for $R=100$ iterations. 
The high accuracy of the DPD based methods together with M-estimators RLASSO and Ad-RLASSO are clearly noticable; in particular the AW-DPD-LASSO succeeds in selecting half of the genes. Additionally, the certainty of the predictions can be evaluated using the MAE between predicted probabilities and the real class labels. A boxplot of the MAE produced in $R=100$ iterations with the different methods is plotted in Figure \ref{fig:boxplotBC2} (top right), where the low error produced by Ad-DPD-LASSO and AW-DPD-LASSO methods stand out.

\subsection{Colon Tumour}
Next we study the  Colon tumour dataset published in \cite{Alon1999}. It contains gene expression levels of 40 tumours and 22 normal colon tissues for 6500 human genes obtained with an Affymetrix olgonucleotide array. A subset of $p=2000$ genes with the highest minimal intensity across the samples was used. The aim is to discriminate between tumour and normal observations.
Additionally, there is biological evidence that the samples T2, T30, T33, T36, T37, N8, N12, N34, N36 may be mislabelled and moreover, almost of all these observations have been flagged as possible outliers by different outlier detection algorithms. Again, we are interested in studying the influence of the outlying observations on the fitted models performance.
We apply a base-10 logarithm transform and scale the data before applying the logistic model. 
Mislabelled observations represent approximately  $15\%$ of the sample size, a proportion critically significant considering the high dimensionality of the data. 
Table \ref{tablecolonleukemia} contains the mean predicted model size and mean accuracy with the unclean and cleaned (after the removal of mislabelled observations) data for $R=100$ fitted models. The number of selected genes with adaptive methods is again the lowest, and the large number of genes selected by the LTS method stands out compared to the remaining methods. However, the hardship of mislabelled observation turns out in a significantly higher accuracy with cleaned data of the robust methods, especially Ad-RLASSO and the proposed Ad-DPD-LASSO and AW-DPD-LASSO: most of the incorrectly classified observations in the training set were actually mislabelled observations. Finally, Figure \ref{fig:boxplotC} (bottom left) shows the boxplot of the MAE between predicted probabilities and real classes belonging with all methods,  where AW-DPD-LASSO with low values of $\alpha = 0.1, 0.3$ stands out for its low errors.


\begin{table}
	\caption{\label{tablecolonleukemia} Model size and accuracy of the different methods with unclean data (training set), cleaned data (after outlier removal) and test data for (a)  for \cite{Alon1999} colon tumour (left) and  (b)  for \cite{golub} Leukemia (right) datasets. }
	\centering
	\fbox{\begin{tabular}{lc}
			\begin{tabular}{rrrr}
				& & \multicolumn{2}{c}{Accuracy ($\%$)} \\
				\cline{3-4}
				& MS &  unclean &  cleaned  \\ 
				\hline
				LASSO & 7.52 & 87.58 & 92.63 \\ 
				Ad-LASSO & 4.00 & 84.53 & 89.19 \\ 
				AW-LASSO & 3.81 & 88.00  & 93.06 \\
				LTS & 19.24 & 91.95 & 96.85 \\ 
				RLASSO & 13.00 & 79.03 & 79.63 \\ 
				AD-RLASSO & 7.00 & 90.32 & 94.44 \\ 
				DPD-LASSO $\alpha = $ 0.1 & 8.09 & 89.60 & 94.61 \\ 
				DPD-LASSO $\alpha = $ 0.3 & 7.87 & 89.27 & 94.44 \\ 
				DPD-LASSO $\alpha = $ 0.5 & 7.82 & 88.95 & 94.13 \\ 
				DPD-LASSO $\alpha = $ 0.7 & 7.91 & 88.56 & 93.94 \\ 
				DPD-LASSO $\alpha = $ 1 & 7.82 & 88.26 & 93.67 \\ 
				Ad-DPD-LASSO $\alpha =  $ 0.1 & 5.25 & 90.24 & 96.17 \\ 
				Ad-DPD-LASSO $\alpha = $ 0.3 & 4.98 & 90.11 & 96.52 \\ 
				Ad-DPD-LASSO $\alpha = $ 0.5 & 4.95 & 90.47 & 96.52 \\ 
				Ad-DPD-LASSO $\alpha = $ 0.7 & 4.95 & 90.47 & 96.52 \\ 
				Ad-DPD-LASSO $\alpha = $ 1 & 4.92 & 90.52 & 96.52 \\ 
				AW-DPD-LASSO $\alpha = $ 0.1 & 5.95 & 95.15 & 98.15 \\ 
				AW-DPD-LASSO $\alpha = $ 0.3 & 6.44 & 97.52 & 99.00 \\ 
				AW-DPD-LASSO $\alpha = $ 0.5 & 4.98 & 90.52 & 96.26 \\ 
				AW-DPD-LASSO $\alpha = $ 0.7 & 5.20 & 91.74 & 96.85 \\ 
				AW-DPD-LASSO $\alpha = $ 1 & 4.92 & 91.90 & 96.37 \\ 
			\end{tabular}
			
			&
			\begin{tabular}{lrr}
				& \multicolumn{2}{c}{Accuracy ($\%$)} \\
				\cline{2-3}
				MS &  train &  test  \\ 
				\hline
				14.34 & 99.58 & 91.41 \\ 
				5.01 & 99.97 & 94.12 \\ 
				8.86 & 97.84 & 92.79 \\
				9.94 & 96.39 & 91.41 \\ 
				4.08 & 100.00 & 91.41 \\ 
				4.00 & 100.00 & 94.12 \\
				13.16 & 100.00 & 94.12 \\ 
				13.68 & 100.00 & 91.41 \\ 
				14.36 & 100.00 & 94.12 \\ 
				15.48 & 100.00 & 91.41 \\ 
				12.57 & 99.87 & 94.12 \\ 
				7.62 & 100.00 & 91.41 \\ 
				6.99 & 99.71 & 94.12 \\ 
				6.59 & 99.71 & 91.41 \\ 
				6.81 & 99.71 & 94.12 \\ 
				7.04 & 99.71 & 91.41 \\ 
				8.64 & 100.00 & 94.12 \\ 
				8.58 & 100.00 & 91.41 \\ 
				8.53& 100.00 & 94.12 \\ 
				8.44 & 100.00 & 91.41 \\ 
				8.54 & 99.71 & 94.12 \\  	
			\end{tabular}
	\end{tabular}
}
\end{table}


\subsection{Leukemia }

Our final example is based on the  leukemia dataset published in \cite{golub}. The dataset contains the expression levels of $6817$ human genes obtained from $n=38$ bone marrow mononuclear cells hybridized to high-density oligonucleotide microarrays (HU6800 chip) produced by Affymetrix. 
The chip actually contains $p=7129$ different probe sets from 27 cases of lymphoid precursors (acute lymphoblastic leukemia, ALL) and 11 cases of myeloid precursors (acute myeloid leukemia, AML). Samples were subjected to a priori quality control standards regarding the amount of labelled RNA and the quality of the scanned microarray image. The aim is the classification of acute leukemias into those arising from ALL or from AML. \cite{golub} also included an independent collection of 34 leukemia samples, consisting on 24 bone marrow and 10 peripheral blood samples from different reference laboratories that used different sample preparation protocols. These independent data can be used to test the quality of the predictors.
The literature does not indicate any biological evidence that the dataset contains mislabelling, though \cite{filzmosermaronna} labelled 296 possibly multivariate outlying genes from the original dataset with $7129$ predictive genes, using both training and test set and there is consensus in the literature that the 66th sample observation is an outlier.
We first apply a preprocessing step to the data as suggested in \cite{golub}, thresholding a floor of
100 and a ceiling of 16000, applying base-2 logarithm and standardization of the data so that the expression
measures for each array have mean 0 and variance 1 across genes. Additionally we filter the genes excluding genes with
$\text{max}/ \text{min}\leq 5$ and $(\text{max}-\text{min}) \leq 500$, where max and min refer respectively to the maximum and minimum expression levels of a particular gene across mRNA samples. Table \ref{tablecolonleukemia} contains the mean model size and mean accuracy with the training and test data produced by the different methods in $R=100$ iterations. The high correct classification rate for all methods with leukemia data indicates a lower contamination, where RLASSO, Ad-RLASSO and DPD-based methods are the most accurate in training and test. Moreover, the proposed methods Ad-DPD-LASSO and AW-DPD-LASSO remain competitive with respect to \textcolor{black}{likelihood-based} based methods, and classify observations with lower MAE (Figure \ref{fig:boxplotL}, bottom right), that is, with more certainty.

\section{Concluding remarks}

In this paper we have presented a robust adaptive method for sparse logistic regression. Based on the DPD loss function and general adaptive weighted penalties, we have proposed a family of estimators depending on a parameter $\alpha \geq 0$ controlling the trade-off between efficiency and robustness. Moreover, these estimators are indeed robust for $\alpha > 0,$ as is shown by the IF analysis, and enjoy oracle properties (consistently perform model selection and parameter estimation) for all $\alpha$ under some necessary assumptions. Besides, we have developed a computational algorithm based on the well known IRLS to compute our proposed estimators and we have compared the performance of our proposed estimators to some existing robust and nonrobust methods under pure data and contaminated data  through a simulation study. Our results show a clear improvement in robustness for our proposals while remaining competitive in efficiency. Finally, we employed our proposed method on four real datasets involving  different types of cancer classification, obtaining promising results in each of the four cases. The logistic regression model for high dimensional data works conveniently to study gene expression profiling, as it performs gene identification and patient classification. Further, measurement errors on gene expression could influence the gene selection, and thus, robust methods are particularly advantageous. 

The competent performance of the proposed methods encourages the extension to more general regression models, as the Poisson regression. Concurrently, robust methodology for high dimensional data based on the DPD could be developed to define robust statistical tests for high dimensional data, an inadequately explored field. Besides, a data-driven study
for the selection of the optimal tuning parameter $\alpha,$ as already indicated, remain in our plans in respect of further extension of this work.  


\section*{Supplementary information}

Code and data are available at \href{https://github.com/MariaJaenada/awDPDlasso}{package awDPDlasso}.

\section*{Acknowledgments}
This research is supported by the Spanish Grants PGC2018-095 194-B-100 and FPU 19/01824.
Research of AG is also partially supported by an INSPIRE Faculty Research Grant from
DST and a research grant no. SRG/2020/000072 from SERB, Government of India, India
\textcolor{black}{We are very grateful to the referees and associate editor for their helpful comments and suggestions}


\newpage
\appendix
\begin{center}
	\textbf{Supplementary material for ``Robust adaptive Lasso in high-dimensional logistic regression with an application to genomic classification of cancer patients.''}
	
	Basu, A.$^{1}$, Ghosh, A.$^{1}$; Jaenada, M.$^{2}$ and Pardo, L.$^{2}$\\$^{1}${\small Indian Statistical Institute, India}\\$^{2}${\small Complutense University of Madrid, Spain}
\end{center}

\section{Calculation for Influence functions }
Along with the notations of Section 2 and 3 of the main paper, for convenience, let us denote $L_\alpha((y,\boldsymbol{x}), \boldsymbol{\beta}) =  \rho_\alpha(\boldsymbol{x}^T\boldsymbol{\beta}, y).$ As discussed in Section 3 of the main paper, the IF of $\widehat{\boldsymbol{\beta}}$ with associated functional $\boldsymbol{T}_{\lambda,\alpha}^{\boldsymbol{\beta}}(G)$ can be computed as
\begin{equation} \label{limIF}
	\text{IF}\left((y_t, \boldsymbol{x}_t), \boldsymbol{T}_{\lambda,\alpha}^{\boldsymbol{\beta}}(G),G\right) = \lim_{m \rightarrow \infty} \text{IF}_{p_m}\left((y_t, \boldsymbol{x}_t), \boldsymbol{T}_{\lambda,\alpha,m}^{\boldsymbol{\beta}}(G), G\right),
\end{equation}
where $\text{IF}_{p_m}$ are the influence functions of estimators penalized with continuous and infinitely differentiably functions in both arguments $p_m(s, t(G))$, for all $m = 1,2,...,$ 
and $p_m(s, t(G))$ converges in the Sobolev space to our adaptive weighted penalty $w\left(|t(G)| \right) |s|$ as $m\rightarrow \infty$.
The associated functional $\boldsymbol{T}_{\lambda,\alpha,m}^{\boldsymbol{\beta}}(G)$ minimizes
\begin{equation}
	Q_{\alpha,\lambda,m}\left(\boldsymbol{\beta}\right) = \int L_{\alpha}\left(  \left(  y,\boldsymbol{x}\right) ;\boldsymbol{\beta}\right)  dG\left(  y,\boldsymbol{x}\right)  + \sum_{j=1}^k p_{m,\lambda}(\beta_j, U_j(G)).
\end{equation}
and it $\text{IF}$ can be calculated as a distributional derivative, since $p_m(\beta_j, U_j(G))$ are differentiable. Equating the derivatives of $Q_{\alpha,\lambda,m}\left(\boldsymbol{\beta}\right)$ to zero we obtain the estimating equations
\begin{equation}\label{estimatingeq} 
	\frac{1+\alpha}{n^{\alpha}}\mathbb{E}_{G}\left[ \boldsymbol{\Psi}_\alpha\left(  \left(  y,\boldsymbol{x}\right) ;\boldsymbol{\beta}\right) \boldsymbol{x}
	\right] + \boldsymbol{P}^\ast_m(\boldsymbol{\beta}, \boldsymbol{U}(G)) = \boldsymbol{0}_{k}
\end{equation}
where $\boldsymbol{\Psi}_\alpha$ is defined in Equation (10) of the main paper, $\boldsymbol{P}^\ast_m(\boldsymbol{\beta}, \boldsymbol{U}(G)) $ is a $k$-vector having $j$-th element as $\frac{\partial p_{m,\lambda}}{\partial \beta_j}(\beta_j, U_j(G))$ and $\boldsymbol{\beta} =\boldsymbol{T}_{\lambda,\alpha,m}^{\boldsymbol{\beta}}(G).$  Let us consider
the degenerate contamination distribution $\Lambda_{(y_t,\boldsymbol{x}_t)}$ at the point $(y_t, \boldsymbol{x}_t).$ Substituting $G$ by $G_\varepsilon = (1-\varepsilon) G + \varepsilon\Lambda_{(y_t,\boldsymbol{x}_t)}$ in (\ref{estimatingeq}) and taking derivative with respect to $\varepsilon$ we obtain,
\begin{align*}
	& \frac{\partial}{\partial \varepsilon}\left[ \frac{1+\alpha}{n^{\alpha}}
	\int \boldsymbol{\Psi}_\alpha\left(  \left(  y,\boldsymbol{x}\right) ;\boldsymbol{\beta}_{\varepsilon}\right) \boldsymbol{x} dG_{\varepsilon}
	+ \boldsymbol{P}^\ast_m(\boldsymbol{\beta}_{\varepsilon}, \boldsymbol{U}(G_{\varepsilon}))  \right] \\
	& =  \frac{1+\alpha}{n^{\alpha}}
	\int \frac{\partial}{\partial \varepsilon} \left[\boldsymbol{\Psi}_\alpha\left(  \left(  y,\boldsymbol{x}\right)  ;\boldsymbol{\beta}_{\varepsilon}\right) \boldsymbol{x}\right] dG_{\varepsilon}\\
	& \hspace{0.5cm} + \frac{1+\alpha}{n^{\alpha}}
	\int  \boldsymbol{\Psi}_\alpha\left(  \left(  y,\boldsymbol{x}\right) ;\boldsymbol{\beta}_{\varepsilon}\right) \boldsymbol{x} d(-G+\Lambda_{(y_t,\boldsymbol{x}_t)})
	+  \frac{\partial}{\partial \varepsilon} \boldsymbol{P}^\ast_m(\boldsymbol{\beta}_{\varepsilon}, \boldsymbol{U}(G_{\varepsilon}))   \\
	& =  \frac{1+\alpha}{n^{\alpha}}
	\int \frac{\partial}{\partial \eta} \left[\boldsymbol{\Psi}_\alpha\left(  \left(  y,\boldsymbol{x}\right) ;\boldsymbol{\beta}_{\varepsilon}\right)\right] \frac{\partial \boldsymbol{\beta}_{\varepsilon} }{\partial \varepsilon} \boldsymbol{x}\boldsymbol{x}^T dG_{\varepsilon}\\
	& \hspace{0.5cm}+ \frac{1+\alpha}{n^{\alpha}}
	\int  \boldsymbol{\Psi}_\alpha\left(  \left(  y,\boldsymbol{x}\right) ;\boldsymbol{\beta}_{\varepsilon}\right) \boldsymbol{x} d(-G+\Lambda_{(y_t,\boldsymbol{x}_t)})\\
	& \hspace{0.5cm} + \boldsymbol{P}^{\ast (1)}_m(\boldsymbol{\beta}_{\varepsilon}, \boldsymbol{U}(G_{\varepsilon})) \frac{\partial \boldsymbol{\beta}_{\varepsilon} }{\partial \varepsilon} + \boldsymbol{P}^{\ast (2)}_m(\boldsymbol{\beta}_{\varepsilon}, \boldsymbol{U}(G_{\varepsilon})) \frac{\partial \boldsymbol{U}(G_{\varepsilon}) }{\partial \varepsilon}   = \boldsymbol{0}_{k}
\end{align*}
where $\boldsymbol{P}^{\ast (l)}_m(\boldsymbol{\beta}_{\varepsilon}, \boldsymbol{U}(G_{\varepsilon}))$ denotes a $k \times k$ diagonal matrix of which $j$-th diagonal entry is the derivative of $\boldsymbol{P}^{\ast}_m(\boldsymbol{\beta}_{\varepsilon}, \boldsymbol{U}(G_{\varepsilon}))$ with respect to the $l$-th argument, $l = 1, 2.$
Evaluating at $\varepsilon = 0,$ we obtain the implicit equations of $\text{IF}_{p_m},$
\begin{align*}
	\frac{1+\alpha}{n^{\alpha}}&
	\int \frac{\partial}{\partial \eta} \left[\boldsymbol{\Psi}_\alpha\left(  \left(  y,\boldsymbol{x}\right) ;\boldsymbol{\beta}\right)\right] \text{IF}_{p_m} \boldsymbol{x}\boldsymbol{x}^T dG_{\varepsilon}
	+ \frac{1+\alpha}{n^{\alpha}}
	\int  \boldsymbol{\Psi}_\alpha\left(  \left(  y,\boldsymbol{x}\right) ;\boldsymbol{\beta}\right) \boldsymbol{x} d(-G+\Lambda_{(y_t,\boldsymbol{x}_t)})\\
	&+ \boldsymbol{P}^{\ast (1)}_m(\boldsymbol{\beta}, \boldsymbol{U}(G)) \text{IF}_{p_m}+ \boldsymbol{P}^{\ast (2)}_m(\boldsymbol{\beta}, \boldsymbol{U}(G)) \text{IF}\left((y_t, \boldsymbol{x}_t), \boldsymbol{U}(G), G\right)  = \boldsymbol{0}_{k}
\end{align*}
where $\text{IF}\left((y_t, \boldsymbol{x}_t), \boldsymbol{U}(G), G\right)$ corresponds to the IF of the functional $\boldsymbol{U}$ associated to the initial estimator $\widetilde{\boldsymbol{\beta}}.$ Using that $ -\frac{1+\alpha}{n^{\alpha}}  \int \boldsymbol{\Psi}_\alpha\left(  \left(  y,\boldsymbol{x}\right) ;\boldsymbol{\beta}\right) \boldsymbol{x} dG = \boldsymbol{P}^{\ast}_m(\boldsymbol{\beta}, \boldsymbol{U}(G)),$ collecting terms and assuming that all relevant integrals exist finitely, 
\begin{equation} \label{IFm}
	\begin{aligned}
		\text{IF}_{p_m}\left((y_t, \boldsymbol{x}_t), \boldsymbol{T}_{m}^{\boldsymbol{\beta}}(G), G\right)\\ = - J_\alpha^\ast(G,\boldsymbol{T}_{m}^{\boldsymbol{\beta}}(G))^{-1} \bigg[ &
		\frac{1+\alpha}{n^{\alpha}}\boldsymbol{\Psi}_\alpha\left(  \left(  y_t,\boldsymbol{x}_t\right) ;\boldsymbol{\beta}\right)\boldsymbol{x}_t  + \boldsymbol{P}^{\ast }_{m,\lambda}(\boldsymbol{\beta},\boldsymbol{U}(G))\\
		&+ \boldsymbol{P}^{\ast (2)}_{m,\lambda}(\boldsymbol{\beta},\boldsymbol{U}(G))\text{IF}\left((y_t, \boldsymbol{x}_t), \boldsymbol{U}(G), G\right)\bigg]
	\end{aligned}
\end{equation}
where the matrix $J_\alpha^\ast(G,\boldsymbol{\beta})$ is given by
\begin{equation}
	J_\alpha^\ast(G,\boldsymbol{\beta}) = J_\alpha(G,\boldsymbol{\beta}) + \boldsymbol{P}^{\ast (1)}_{m,\lambda}(\boldsymbol{\beta},\boldsymbol{U}(G))
\end{equation}
with
\[ 
J_\alpha(G,\boldsymbol{\beta}) = \mathbb{E}_G\left[ \frac{\partial^2 L_{\alpha}\left(  \left(  y,\boldsymbol{x}\right) ;\boldsymbol{\beta}\right)}{\partial \boldsymbol{\beta}\boldsymbol{\beta}^T} \right]
\]
and
\begin{align*}
	& \frac{\partial^2 L_{\alpha}\left( \left(  y,\boldsymbol{x}\right) ;\boldsymbol{\beta}\right)}{\partial \boldsymbol{\beta}\boldsymbol{\beta}^T}\\ &=  \frac{\alpha+1}{n^\alpha}\left( 1 + e^{\boldsymbol{x}^T\boldsymbol{\beta}}\right)^{-(\alpha+2)}
	\bigg[\frac{e^{(\alpha+1)\boldsymbol{x}^T\boldsymbol{\beta}}}{1+e^{\boldsymbol{x}^T\boldsymbol{\beta}}}\left[1+\alpha-e^{\boldsymbol{x}^T\boldsymbol{\beta}} \right] + \frac{e^{2\boldsymbol{x}^T\boldsymbol{\beta}}}{1+e^{\boldsymbol{x}^T\boldsymbol{\beta}}}\left[2-\alpha e^{\boldsymbol{x}^T\boldsymbol{\beta}}\right]\\
	& \hspace{0.5cm} -  y\left[\alpha e^{\alpha\boldsymbol{x}^T\boldsymbol{\beta}} - e^{(\alpha+1)\boldsymbol{x}^T\boldsymbol{\beta}}+ e^{\boldsymbol{x}^T\boldsymbol{\beta}} -\alpha e^{2\boldsymbol{x}^T\boldsymbol{\beta}}\right] \boldsymbol{x}\boldsymbol{x}^T
	\bigg].
\end{align*}
Then, using that $\mathbb{E}_{(\boldsymbol{X},Y)} = \mathbb{E}_{\boldsymbol{X}}\mathbb{E}_{Y/\boldsymbol{X}}$ the above expectation is given by
\begin{equation}
	\begin{aligned}
		J_\alpha(G,\boldsymbol{\beta}) = \frac{\alpha+1}{n^\alpha}\mathbb{E}_{\boldsymbol{X}} \bigg[ \left( 1 + e^{\boldsymbol{x}^T\boldsymbol{\beta}}\right)^{-(\alpha+3)}
		\bigg[ e^{(\alpha+1)\boldsymbol{x}^T\boldsymbol{\beta}} - e^{2\boldsymbol{x}^T\boldsymbol{\beta}}
		\bigg] \boldsymbol{x}\boldsymbol{x}^T\bigg].
	\end{aligned}
\end{equation}
Based on (\ref{limIF}), it suffices to select a particular choice of the sequence of continuous and infitely differentiable $p_{m,\lambda}$ and take limits as $m\rightarrow \infty$ to obtain the expression of $\text{IF}\left((y_t, \boldsymbol{x}_t), \boldsymbol{T}_{\lambda,\alpha}^{\boldsymbol{\beta}}(G),G\right).$ Following \cite{gm}, we define $p_{m,\lambda}(s,U(G)) = \lambda w (h_m(U(G)))h_m(s)$ where  $h_m$ is given by
$h_m(s) = \frac{2}{m}\log(e^{sm}+1)-s \xrightarrow{m\rightarrow\infty} |s|.$ Taking limits in (\ref{IFm}) for this particular choice of $p_{m,\lambda}$ we straightforward obtain the expression given in Theorem 1 of the main paper.

\section{Important Lemmas required}

For the proof of Theorem 3 and 5 we need three aid results, given in Lemma 12 and 15 of Supplementary material of \cite{gm} and Lemma 4 of Supplementary material of \cite{fan2014} respectively. We only present here their formulation for the sake of completeness.

\begin{lemma}[Lemma 12,  \cite{gm}, Supplementary material]\label{LEM:L1}
	Consider $\rho_{\alpha}(\boldsymbol{x}_i^T\boldsymbol{\beta}, y_i)$ the function defined in Equation (20) of the main paper and define 
	$$\mathcal{B}_0(M) =\left\{ \boldsymbol{\beta}\in\mathbb{R}^p : \parallel\boldsymbol{\beta} - \boldsymbol{\beta}_0 \parallel_2\leq M, 
	Supp(\boldsymbol{\beta}) \subseteq Supp(\boldsymbol{\beta}_0) = S_0 \right\}$$ 
	and 
	\begin{eqnarray}
		Z_n(M) = \sup\limits_{\boldsymbol{\beta}\in \mathcal{B}_0(M)}  \bigg\rvert &\sum_{i=1}^{n} \left( \rho_{\alpha}(\boldsymbol{x}_i^T\boldsymbol{\beta}, y_i) - 
		\rho_{\alpha}(\boldsymbol{x}_i^T\boldsymbol{\beta}_0, y_i) \right) \nonumber \\
		&-E\left(\sum_{i=1}^{n} \left(\rho_{\alpha}(\boldsymbol{x}_i^T\boldsymbol{\beta}, y_i) -  \rho_{\alpha}(\boldsymbol{x}_i^T\boldsymbol{\beta}_0, y_i) \right)\right)\bigg\rvert.
		\label{EQ:Z_nM}
	\end{eqnarray}
	Under Assumptions (A1)--(A2), it holds for any $t>0$
	\begin{eqnarray}
		P\left(Z_n(M) \geq 4ML\sqrt{s/n} + t \right) \leq e^{-\frac{nc_0t^2}{8M^2L^2}},
	\end{eqnarray}
	where $M$ is the radius of $\mathcal{B}_0(M)$, $n$ is the sample size, $s$ is the number of true non-zero coefficients, $L$ is the common Lipschitz constant defined in (A1) and $c_0$ is defined in (A2).
\end{lemma}

\begin{lemma}[Lemma 15,  \cite{gm}, Supplementary material]\label{LEM:L3}
	Let define $\eta_{i0}=\boldsymbol{x}_{1i}^T\boldsymbol{\beta}_{10}$ for $i=1, \ldots, n$ and  
	$\boldsymbol{\theta} = \boldsymbol{V}_n^{-1}(\boldsymbol{\beta}_1 - \boldsymbol{\beta}_{10}).$ Also define 
	$$R_{n,i}(\boldsymbol{\theta}) = \rho_{\alpha}(\eta_{i0} + \boldsymbol{Z}_{n,i}^T\boldsymbol{\theta}, y_i) - \rho_{\alpha}(\eta_{i0}, y_i)
	+ (1+\alpha)\boldsymbol{\Psi}_{\alpha}(\eta_{i0},y_i) \boldsymbol{Z}_{n,i}^T\boldsymbol{\theta},$$ 
	for each $i,$
	$R_n(\boldsymbol{\theta}) = \sum_{i=1}^n R_{n,i}(\boldsymbol{\theta})$ and $r_n(\boldsymbol{\theta}) = R_n(\boldsymbol{\theta}) - E[R_n(\boldsymbol{\theta})].$
	Consider $\boldsymbol{\theta}$ over the convex open set 
	$B_0(n) = \left\{\boldsymbol{\theta}\in \mathbb{R}^s : \parallel\boldsymbol{\theta}\parallel_2 <c^* \sqrt{s} \right\}$ 
	for some constant $c^*>0$ independent of $s$. Then, under Assumptions of Theorem 5, we have, for any $\epsilon>0$ and $\boldsymbol{\theta}\in B_0(n)$, 
	$$
	P\left(\rvert r_n(\boldsymbol{\theta})\rvert\geq \epsilon)\right) \leq \exp\left(-C\epsilon b_ns^2(\log s)\right), 
	$$
	where $b_n$ is some diverging sequence such that $b_ns^{7/2}(\log s) \max_i \parallel\boldsymbol{Z}_{n,i}\parallel_2 \rightarrow 0$ and
	$C>0$ is some constant.
\end{lemma}

\begin{lemma}[Lemma 4, \cite{fan2014}, Supplementary material]\label{LEM:L4}
	Let $h(\boldsymbol{\theta})$ be a positive function defined on the convex open set $B_0(n)$,
	and $\left\{ h_n(\boldsymbol{\theta}): \boldsymbol{\theta}\in B_0(n) \right\}$ be a sequence of random convex functions.
	Suppose that there exists a diverging sequence $b_n$ such that, for every $\boldsymbol{\theta}\in B_0(n)$ and for all $\epsilon>0$, 
	$$
	P\left(\rvert h_n(\boldsymbol{\theta}) - h(\boldsymbol{\theta})\rvert \geq \epsilon \right)
	\leq c_4 \exp\left[ - c_5 \epsilon b_n s^2(\log s)\right],
	$$ 
	for some constants $c_4, c_5 >0$. Further, assume that there exists a constant $c_6>0$ such that 
	$
	\rvert h(\boldsymbol{\theta}_1) - h(\boldsymbol{\theta}_2)\rvert  \leq c_6 s \parallel\boldsymbol{\theta}_1 - \boldsymbol{\theta}_2\parallel_{\infty},
	$
	for any $\boldsymbol{\theta}_1, \boldsymbol{\theta}_2 \in B_0(n)$.
	Then, we have 
	$$
	\sup_{K_s}\left\rvert  h_n(\boldsymbol{\theta}) - h(\boldsymbol{\theta})\right\rvert  = o_P(1),
	$$
	where $K_s$ is any compact set in $\mathbb{R}^s$ defined as 
	$K_s =\left\{ \boldsymbol{\theta}\in \mathbb{R}^s : \parallel\boldsymbol{\theta}\parallel_2 \leq c_7\sqrt{s} \right\} \subset B_0(n)$ 
	for some $c_7 \in (0, c^*)$.
\end{lemma}

\section{Proof of the Main Theorems }
In the following proofs we use the same notation as in the main paper. 

\subsection*{Proof of Theorem 2}

Let us consider the subspace of all vectors $\boldsymbol{\beta}$ in a neighbourhood of the true parameter value $\boldsymbol{\beta}_0$ of radius $M >0,$  such that all non-zero elements of $\boldsymbol{\beta}$ belongs to $\mathcal{S}_0$,
$$\mathcal{B}_0(M) =\left\{ \boldsymbol{\beta}\in\mathbb{R}^p : \parallel\boldsymbol{\beta} - \boldsymbol{\beta}_0 \parallel_2\leq M, 
Supp(\boldsymbol{\beta}) \subseteq Supp(\boldsymbol{\beta}_0) = S_0 \right\}.$$ 
The previously defined subspace is clearly non empty. Recall that $\boldsymbol{\beta} = (\boldsymbol{\beta}_1^T,\boldsymbol{0})^T$, where $\boldsymbol{\beta}_1$ contains the first $s$ elements of the vector $\boldsymbol{\beta}$. Note that $\boldsymbol{\beta}_{01}$ has not zero elements, while $\boldsymbol{\beta}_1$ may have it, for all $\boldsymbol{\beta} \in \mathcal{B}_0(M).$ We will first bound the difference between $\boldsymbol{\beta}_1$ and $\boldsymbol{\beta}_{01}$ for any $\boldsymbol{\beta} \in \mathcal{B}_0(M).$
A Taylor series expansion of $\sum_{i=1}^{n} \rho_{\alpha}(\boldsymbol{x}_i^T\boldsymbol{\beta}, y_i)$ at any 
$\boldsymbol{\beta}=(\boldsymbol{\beta}_1^T, \boldsymbol{0}^T)^T\in\mathcal{B}_0(M)$ around $\boldsymbol{\beta}_0$, 
along with Assumption (A4), yields 
\begin{equation}
	\begin{aligned}
		&E\left[\sum_{i=1}^{n} \left(\rho_{\alpha}(\boldsymbol{x}_i^T\boldsymbol{\beta}, y_i) - \rho_{\alpha}(\boldsymbol{x}_i^T\boldsymbol{\beta}_0, y_i)\right)\right]\\
		&= \frac{1}{2n}(\boldsymbol{\beta}_1 - \boldsymbol{\beta}_{10})^TE[\boldsymbol{X}_1^T\boldsymbol{H}_{\gamma}^{(2)}(\boldsymbol{\beta}_0)\boldsymbol{X}_1] 
		(\boldsymbol{\beta}_1 - \boldsymbol{\beta}_{10})
		+ o(1),
	\end{aligned} 
	\label{EQ:A.1}
\end{equation}
since the first order derivative of $E[\sum_{i=1}^{n} \rho_{\alpha}(\boldsymbol{x}_i^T\boldsymbol{\beta}, y_i)]$ is zero at $\boldsymbol{\beta}=\boldsymbol{\beta}_0$.
The first term of the right hand size can be bounded bellow using Assumptions (A2) and (A3), giving 
$$
\frac{1}{n}(\boldsymbol{\beta}_1 - \boldsymbol{\beta}_{10})^TE[\boldsymbol{X}_1^T\boldsymbol{H}^{(2)}(\boldsymbol{\beta}_0)\boldsymbol{X}_1] 
(\boldsymbol{\beta}_1 - \boldsymbol{\beta}_{10}) \geq c_0c_1\parallel\boldsymbol{\beta}_1 - \boldsymbol{\beta}_{10}\parallel_2^2,
$$
for sufficiently large $n$, which finally implies
\begin{equation}
	E\left[\sum_{i=1}^{n} \left(\rho_{\alpha}(\boldsymbol{x}_i^T\boldsymbol{\beta}, y_i) - \rho_{\alpha}(\boldsymbol{x}_i^T\boldsymbol{\beta}_0, y_i)\right)\right]
	\geq \frac{1}{2} c_0c_1\parallel\boldsymbol{\beta}_1 - \boldsymbol{\beta}_{10}\parallel_2^2. 
	\label{EQ:ThP.1}
\end{equation}
The restricted oracle estimator $\widehat{\boldsymbol{\beta}}$ may not be in $\mathcal{B}_0(M)$. However, we can define $\widetilde{\boldsymbol{\beta}}_1 = u \widehat{\boldsymbol{\beta}}_1 + (1-u) \boldsymbol{\beta}_{10}$
with $u = M/(M+\parallel\widehat{\boldsymbol{\beta}}_1 - \boldsymbol{\beta}_{10}\parallel_2)$ so that 
$\widetilde{\boldsymbol{\beta}} = (\widetilde{\boldsymbol{\beta}}_1^T, \boldsymbol{0}^T)^T \in \mathcal{B}_0(M)$ and satisfies (\ref{EQ:ThP.1}). We now derive a bound for the left hand size expectation in (\ref{EQ:ThP.1})  with probability tending to 1. Let us consider, as a function of $M$, the quantity defined in Lemma \ref{LEM:L1}, 
\begin{equation*}
	\begin{aligned}
		Z_n(M) = \sup\limits_{\boldsymbol{\beta}\in \mathcal{B}_0(M)} \bigg\rvert & \sum_{i=1}^{n} \left( \rho_{\alpha}(\boldsymbol{x}_i^T\boldsymbol{\beta}, y_i) - 
		\rho_{\alpha}(\boldsymbol{x}_i^T\boldsymbol{\beta}_0, y_i) \right)\\
		&-E\left(\sum_{i=1}^{n} \left(\rho_{\alpha}(\boldsymbol{x}_i^T\boldsymbol{\beta}, y_i) -  \rho_{\alpha}(\boldsymbol{x}_i^T\boldsymbol{\beta}_0, y_i) \right)\right)\bigg\rvert 
	\end{aligned}
\end{equation*}
Since the oracle estimator is a minimum of the convex objective function $Q_{n,\alpha, \lambda},$  and by the definition of $\widetilde{\boldsymbol{\beta}},$ we have that $Q_{n,\alpha, \lambda}(\widetilde{\boldsymbol{\beta}}) \leq Q_{n,\alpha, \lambda}(\boldsymbol{\beta}_0).$ Therefore,
$$\sum_{i=1}^n \left( \rho_\alpha(\boldsymbol{x}_i^T\boldsymbol{\beta}, y_i) - 
\rho_\alpha(\boldsymbol{x}_i^T\boldsymbol{\beta}_0, y_i) \right) \leq \lambda_n \sum_{j=1}^k \left(\rvert \beta_{0j}\rvert -\rvert \widetilde{\beta}_j\rvert \right) \leq \lambda_n \sum_{j=1}^k \big\rvert \widetilde{\beta}_j - \beta_{0j}\big\rvert  $$
and hence, using the definition of $Z_n(M)$, we have that
\begin{eqnarray}
	E\left[ \sum_{i=1}^n \left( \rho_\alpha(\boldsymbol{x}_i^T\boldsymbol{\beta}, y_i) - 
	\rho_\alpha(\boldsymbol{x}_i^T\boldsymbol{\beta}_0, y_i) \right)\right]
	&\leq& Z_n(M) + \sum_{i=1}^n \left( \rho_\alpha(\boldsymbol{x}_i^T\boldsymbol{\beta}, y_i) - 
	\rho_\alpha(\boldsymbol{x}_i^T\boldsymbol{\beta}_0, y_i) \right)
	\nonumber\\
	&\leq& Z_n(M) + \lambda_n \sum\limits_{j=1}^{s} w_{j}\left\vert \widetilde{\beta}_{j} - \beta_{j0}\right\vert
	\nonumber\\
	&\leq& Z_n(M) + \lambda_n \parallel\boldsymbol{w}_0\parallel_2 \parallel\widetilde{\boldsymbol{\beta}}_1 - \boldsymbol{\beta}_{01}\parallel_2\\ \nonumber
	&\leq& Z_n(M) + \lambda_n \parallel\boldsymbol{w}_0\parallel_2 M,\nonumber
\end{eqnarray}
where the second last inequality holds by Cauchy-Schwarz inequality.
We finally define the event  $\mathcal{E}_n = \left\{Z_n(M) \leq 2MLn^{-1/2}\sqrt{s\log n} \right\}$ under which the last inequality implies
\begin{eqnarray}
	E\left[ \sum_{i=1}^n \left( \rho_\alpha(\boldsymbol{x}_i^T\boldsymbol{\beta}, y_i) - 
	\rho_\alpha(\boldsymbol{x}_i^T\boldsymbol{\beta}_0, y_i) \right)\right]
	\leq 2MLn^{-1/2}\sqrt{s\log n} + \lambda_n \parallel\boldsymbol{w}_0\parallel_2 M,
	\label{EQ:ThP.2}
\end{eqnarray}
If we take  $M=L^{-1}\left[2\sqrt{s/n} + \lambda_n \parallel\boldsymbol{w}_0\parallel_2\right],$ so that it satisfies $M=o(\kappa_n^{-1}s^{-1/2})$
by Assumption (A2) and the fact that $\lambda_n\parallel\boldsymbol{w}_0\parallel_2\sqrt{s}\kappa_n \rightarrow 0$, we can combine inequalities (\ref{EQ:ThP.1}) for $\boldsymbol{\beta}_1 = \widetilde{\boldsymbol{\beta}}_1$ 
and (\ref{EQ:ThP.2}), on the event $\mathcal{E}_n$, to obtain that 
\begin{eqnarray}
	\frac{1}{2} c_0c_1\parallel\widetilde{\boldsymbol{\beta}}_1 - \boldsymbol{\beta}_{10}\parallel_2^2
	&\leq& \left(2n^{-1/2}\sqrt{s\log n} + \lambda_n \parallel\boldsymbol{w}_0\parallel_2\right) ML
	\nonumber\\
	&\leq& \left(2n^{-1/2}\sqrt{s\log n} + \lambda_n \parallel\boldsymbol{w}_0\parallel_2\right) \left(2\sqrt{s/n} + \lambda_n \parallel\boldsymbol{w}_0\parallel_2\right),
	\nonumber
\end{eqnarray}
and hence 
\begin{eqnarray}
	\parallel\widetilde{\boldsymbol{\beta}}_1 - \boldsymbol{\beta}_{10}\parallel_2
	&\leq& O\left(\sqrt{s(\log n)/n} + \lambda_n \parallel\boldsymbol{w}_0\parallel_2\right).
\end{eqnarray}
But $\parallel\widetilde{\boldsymbol{\beta}}_1 - \boldsymbol{\beta}_{10}\parallel_2 = \frac{M \parallel\widehat{\boldsymbol{\beta}}^o_1 - \boldsymbol{\beta}_{10}\parallel_2}{
	M + \parallel\widehat{\boldsymbol{\beta}}^o_1 - \boldsymbol{\beta}_{10}\parallel_2}\leq M/2$, and therefore
$$
\parallel\widehat{\boldsymbol{\beta}}^o_1 - \boldsymbol{\beta}_{10}\parallel_2 \leq M 
= O\left(2\sqrt{s/n} + \lambda_n \parallel\boldsymbol{w}_0\parallel_2\right).
$$
on the event $\mathcal{E}_n$.  To complete the demonstration, it only remains to show that $\mathcal{E}_n$ is satisfied with probability tending to 1.  That condition is equivalent to that the function $Z_n(M)$ can be bounded with probability tending to $1,$ what is straightforward using lemma \ref{LEM:L1}.
Without any loss of generality take $L=(1+\alpha)L/\alpha > 1$ so that, by applying Lemma \ref{LEM:L1} with $t=ML\sqrt{s(\log n)/n}$, 
we get for large enough $n$ (satisfying $\log n \geq 4$)
\begin{eqnarray}
	P(\mathcal{E}_n^c) &=& P\left(Z_n(M) > 2MLn^{-1/2}\sqrt{s\log n}\right)
	\nonumber\\
	&\leq& P\left(Z_n(M) > 2ML \sqrt{s/n} + t\right) \leq \exp(-{c_0s(\log n)}/{8}).
	\nonumber
\end{eqnarray}
The first inequality follows because $2ML\sqrt{s/n} + t < 2MLn^{-1/2}\sqrt{s\log n}$.
Hence, we get
$$
P(\mathcal{E}_n) \geq 1 - \exp(-{c_0s(\log n)}/{8}) = 1 - n^{-c_0s/8}.  
$$
The second part of the Theorem is straightforward.

\subsection*{Proof of Theorem 3}
The proof follows exactly in the same manner as in the proof of Theorem 2 of Fan et al. \cite{fan2014}, as detailed in the supplementary material of \cite{gm}.

\subsection*{Proof of Theorem 4}

We follow the same line as  the proof of Theorem 3 of \cite{fan2014} and Theorem 8 of \cite{gm}. Let us consider the quantities defined in Lemma \ref{LEM:L3} and 
define the objective function restricted to the subspace of all vectors $\boldsymbol{\beta}$ with null $k-s$ last elements,
$Q_{n,\gamma}^\ast(\boldsymbol{\theta})  = n \left[Q_{n,\gamma,\lambda_n}(\boldsymbol{\beta}_1, \boldsymbol{0}) - Q_{n,\gamma, \lambda_n}(\boldsymbol{\beta}_{01}, \boldsymbol{0})\right].$
Theorem 4 of the main paper claims the oracle estimator, $\widehat{\boldsymbol{\beta}} = (\widehat{\boldsymbol{\beta}}_1^T, \widehat{\boldsymbol{\beta}}_2^T)^T$ satisfies $\widehat{\boldsymbol{\beta}}_2 = \boldsymbol{0}_{k-s}$ with probability tending to 1, and therefore $\boldsymbol{\beta}_1=\widehat{\boldsymbol{\beta}}_1$ is the minimizer of the restricted objective function. That is, $Q_{n,\gamma}^\ast(\boldsymbol{\theta})$ reaches its minimum at $\widehat{\boldsymbol{\theta}}_n 
= \boldsymbol{V}_n^{-1}\left(\widehat{\boldsymbol{\beta}}_1 - \boldsymbol{\beta}_{01}\right).$

We now decompose $Q_{n,\gamma}^\ast(\boldsymbol{\theta})$ into  its mean and centralized stochastic component as 
$
Q_{n,\gamma}^\ast(\boldsymbol{\theta}) = M_n(\boldsymbol{\theta}) + T_n(\boldsymbol{\theta}),\nonumber
$
where 
\begin{equation}
	\begin{aligned}
		M_n(\boldsymbol{\theta}) &=E[Q_{n,\gamma}^\ast(\boldsymbol{\theta})] 
		= nE\left[\sum_{i=1}^{n} \left(\rho_{\alpha}(\boldsymbol{x}_i^T(\boldsymbol{\beta}_1,\boldsymbol{0}), y_i) - \rho_{\alpha}(\boldsymbol{x}_i^T(\boldsymbol{\beta}_{01},\boldsymbol{0}), y_i)\right)\right] \\
		&~~~~~~~~~~~~~~~~~~~~~~~~~~+ n\lambda_n\sum\limits_{j=1}^{p} w_{j}\left(\left\vert \beta_{0j} + (\boldsymbol{V}_n\boldsymbol{\theta})_j\right\vert 
		- \left\vert \beta_{0j}\right\vert\right),
		\nonumber\\
		T_n (\boldsymbol{\theta}) &=  Q_{n,\gamma}^\ast(\boldsymbol{\theta}) - E[Q_{n,\gamma}^\ast(\boldsymbol{\theta})]
		= r_n(\boldsymbol{\theta}) - \boldsymbol{H}_{\gamma}^{(1)}(\boldsymbol{\theta}_0)\boldsymbol{Z}_n\boldsymbol{\theta}.
	\end{aligned}
\end{equation}
Here, we have used that $\boldsymbol{Z}_n = \boldsymbol{X}_1\boldsymbol{V}_n$ and  $E[\boldsymbol{H}_{\gamma}^{(1)}(\boldsymbol{\theta}_0)\boldsymbol{Z}_n\boldsymbol{\theta}]=\boldsymbol{0}.$ We first study the mean stochastic component. Taking expectation in a Taylor series expansion of $\sum_{i=1}^{n}\rho_{\alpha}(\boldsymbol{x}_i^T(\boldsymbol{\beta}_1,\boldsymbol{0}), y_i)$ around $\boldsymbol{\beta}_0$ gives 
\begin{eqnarray}
	&nE\left[\sum_{i=1}^{n} \left(\rho_{\alpha}(\boldsymbol{x}_i^T(\boldsymbol{\beta}_1,\boldsymbol{0}), y_i) - \rho_{\alpha}(\boldsymbol{x}_i^T(\boldsymbol{\beta}_{01},\boldsymbol{0}), y_i)\right)\right]  \nonumber \\
	&= \frac{1}{2}  \boldsymbol{\theta}^T(\boldsymbol{Z}_n^TE[\boldsymbol{H}_{\gamma}^{(2)}(\boldsymbol{\beta}_0)]\boldsymbol{Z}_n)\boldsymbol{\theta}  +o(1)\\
	&= \frac{1}{2}\parallel\boldsymbol{\theta}\parallel_2^2 + o(1).
	\nonumber
\end{eqnarray}
For the second term, we are showing that $sign(\boldsymbol{\beta}_{01}+ \boldsymbol{V}_n\boldsymbol{\theta}) = sign(\boldsymbol{\beta}_{01})$ at $B_0(n)$ and hence we can omit the absolute value, obtaining 
$$
\sum\limits_{j=1}^{p} w_{j}\left(\left\vert \beta_{0j} + (\boldsymbol{V}_n\boldsymbol{\theta})_j\right\vert 
- \left\vert \beta_{0j}\right\vert\right) = \widetilde{\boldsymbol{w}_0}^T\boldsymbol{V}_n\boldsymbol{\theta}.
$$
But by Assumptions (A2), (A3) and (A6) implies
$$
\parallel\boldsymbol{V}_n\boldsymbol{\theta}\parallel_{\infty} \leq \parallel\boldsymbol{V}_n\boldsymbol{\theta}\parallel_2
\leq C n^{-1/2} \parallel\boldsymbol{\theta}\parallel_2 = o\left(\min\limits_{1\leq j \leq s}\rvert \beta_{j0}\rvert \right), 
$$
and then the previous condition is fulfilled. Thus,
\begin{eqnarray}
	Q_{n,\gamma}^\ast(\boldsymbol{\theta}) 
	&=& \frac{1}{2}\parallel\boldsymbol{\theta}\parallel_2^2 + n \lambda_n\widetilde{\boldsymbol{w}_0}^T\boldsymbol{V}_n\boldsymbol{\theta} 
	- \boldsymbol{H}_{\gamma}^{(1)}(\boldsymbol{\theta}_0)\boldsymbol{Z}_n\boldsymbol{\theta}  + r_n(\boldsymbol{\theta}) + o(1)
	\nonumber\\
	&=& \frac{1}{2} \parallel\boldsymbol{\theta} - \boldsymbol{\eta}_n\parallel_2^2 - \frac{1}{2}\parallel\boldsymbol{\eta}_n\parallel_2^2 + r_n(\boldsymbol{\theta}) + o(1),
	\label{EQ:Ct.2}
\end{eqnarray} 
with $\boldsymbol{\eta}_n = 
\left[\boldsymbol{H}_{\gamma}^{(1)}(\boldsymbol{\theta}_0)\boldsymbol{Z}_n - n \lambda_n\boldsymbol{V}_n\widetilde{\boldsymbol{w}_0}\right]$. Now, by 
by the central limit theorem we have 
$\boldsymbol{u}^T[\boldsymbol{Z}_n^T\boldsymbol{\Omega}_n\boldsymbol{Z}_n]^{-1/2} \boldsymbol{H}_{\gamma}^{(1)}(\boldsymbol{\theta}_0)\boldsymbol{Z}_n$
asymptotically follows a standard normal distribution for any $\boldsymbol{u}\in \mathbb{R}^s$ with $\boldsymbol{u}^T\boldsymbol{u}=1$, 
and hence 
\begin{eqnarray}
	\boldsymbol{u}^T[\boldsymbol{Z}_n^T\boldsymbol{\Omega}_n\boldsymbol{Z}_n]^{-1/2} \left[ \boldsymbol{\eta}_n 
	+ n\lambda_n\boldsymbol{V}_n\widetilde{\boldsymbol{w}_0} \right]
	\mathop{\rightarrow}^{\mathcal{L}} N(0,1).
	\label{EQ:Ct.3}
\end{eqnarray}
This ensures $\boldsymbol{\eta}_n $ is asymptotically normal. Moreover, $\boldsymbol{\eta}_n$ is bounded in $\ell_2$-norm since
\begin{eqnarray}
	\parallel\boldsymbol{\eta}_n\parallel_2 &\leq&  
	\parallel\boldsymbol{H}_{\gamma}^{(1)}(\boldsymbol{\theta}_0)\boldsymbol{Z}_n\parallel_2 + n \lambda_n\parallel\boldsymbol{V}_n\widetilde{\boldsymbol{w}_0}\parallel_2
	\nonumber\\
	&\leq& O_P(\sqrt{s}) + C\lambda_n\sqrt{n}\parallel\widetilde{\boldsymbol{w}_0}\parallel_2
	~~~~~~\mbox{[Using Assumptions (A2)-(A3)]}
	\nonumber\\
	&=& C\sqrt{s}\left[1 + O_P(1)\right],
	~~~~~~~~~~~~\mbox{[Using Assumption (A6)].}
	\label{EQ:Ct.3a}
\end{eqnarray}
And therefore $\boldsymbol{\eta}_n \in B_0(n) $ for large enough $c^\ast \gg C.$
But we are interested in showing the asymptotic normality of $\widehat{\boldsymbol{\theta}}_n 
+ n\lambda_n\boldsymbol{V}_n\widetilde{\boldsymbol{w}_0}$, so we will show that  $\boldsymbol{\eta}_n$ is close enough to  $\widehat{\boldsymbol{\theta}}_n$  with probability tending to one. 
For this purpose, using Lemma \ref{LEM:L3}, 
we get a sequence $b_n \rightarrow \infty$ such that, for any $\epsilon>0$, 
$$
P\left(\rvert r_n(\boldsymbol{\theta})\rvert  \geq \epsilon \right) \leq \exp\left[-C\epsilon b_n s (\log s)\right].
$$
To apply Lemma \ref{LEM:L4}, we note that $r_n(\boldsymbol{\theta})$ can be written as 
$r_n(\boldsymbol{\theta}) = h_n(\boldsymbol{\theta}) - h(\boldsymbol{\theta}) +o(1)$ for 
$$
h_n(\boldsymbol{\theta}) = Q_{n,\gamma}^\ast(\boldsymbol{\theta}) - n \lambda_n\widetilde{\boldsymbol{w}_0}^T\boldsymbol{V}_n\boldsymbol{\theta} 
+ \boldsymbol{H}_{\gamma}^{(1)}(\boldsymbol{\theta}_0)\boldsymbol{Z}_n\boldsymbol{\theta},
~~~~\mbox{ and }~~
h(\boldsymbol{\theta}) = \parallel\boldsymbol{\theta}\parallel_2^2.
$$
By definition, these functions $h_n(\boldsymbol{\theta})$ and $h(\boldsymbol{\theta})$ are convex on $B_0(n)$ and, by Assumption (A3)
we have for any $\boldsymbol{\theta}_1, \boldsymbol{\theta}_2 \in B_0(n)$, 
\begin{eqnarray}
	\left\rvert h(\boldsymbol{\theta}_1) - h(\boldsymbol{\theta}_2)\right\rvert  &=&  \parallel\boldsymbol{\theta}_1\parallel_2^2- \parallel\boldsymbol{\theta}_2\parallel_2^2 =
	\left\rvert (\boldsymbol{\theta}_1+ \boldsymbol{\theta}_2)^T(\boldsymbol{\theta}_1 -  \boldsymbol{\theta}_2)\right\rvert 
	\nonumber\\
	&\leq& \parallel\boldsymbol{\theta}_1+ \boldsymbol{\theta}_2\parallel_2 \parallel\boldsymbol{\theta}_1 - \boldsymbol{\theta}_2\parallel_2
	\leq C s \parallel\boldsymbol{\theta}_1 -  \boldsymbol{\theta}_2\parallel_{\infty}.\nonumber
\end{eqnarray}
Thus, all the conditions of Lemma \ref{LEM:L4} are satisfies and we have 
$
\sup_{K_s}\left\rvert  r_n(\boldsymbol{\theta})\right\rvert  = o_P(1),
$
for any compact set  $K_s =\left\{ \boldsymbol{\theta}\in \mathbb{R}^s : \parallel\boldsymbol{\theta}\parallel_2 \leq c_7\sqrt{s} \right\} \subset B_0(n)$ 
with $c_7 \in (0, c^*)$.

Since $\boldsymbol{\eta}_n \in B_0(n)$, we can choose $c_7$ large enough such that, for each $s$, the corresponding set $K_s$ cover the ball 
$B_1(n)$ centered at $\boldsymbol{\eta}_n$ and radius $\epsilon$, an arbitrary fixed positive constant, with probability tending to one.
Then, we get 
\begin{eqnarray}
	\Delta_n = \sup\limits_{\boldsymbol{\theta}\in B_1(n)} \rvert r_n(\boldsymbol{\theta})\rvert  
	\leq \sup_{K_s}\left\rvert  r_n(\boldsymbol{\theta})\right\rvert  = o_P(1).
	\label{EQ:Ct.4}
\end{eqnarray}

Finally, we show that indeed the minimizer $\widehat{\boldsymbol{\theta}}_n$ of $Q_{n,\gamma}^\ast(\boldsymbol{\theta})$ must
lie within the ball $B_1(n)$ with probability tending to one, studying the behaviour of $Q_{n,\gamma}^\ast(\boldsymbol{\theta})$
outside $B_1(n)$.
Any vector outside $B_1(n)$ can be written as $\boldsymbol{\theta} = \boldsymbol{\eta}_n + a \boldsymbol{e} \in \mathbb{R}^s$ 
for $\boldsymbol{e}\in \mathbb{R}^s$ a unit vector and $a$ a constant satisfying $a>\epsilon$.
Let $\boldsymbol{\theta}^\ast = \boldsymbol{\eta}_n + \epsilon \boldsymbol{e} = (1 - \epsilon/a) \boldsymbol{\eta}_n + (\epsilon/a)\boldsymbol{\theta}$ be the boundary point of $B_1(n)$ that lies on the line segment joining $\boldsymbol{\eta}_n$ and $\boldsymbol{\theta}.$
Using the convexity of $Q_{n,\gamma}^\ast(\boldsymbol{\theta})$, along with (\ref{EQ:Ct.2}), we get \\
\begin{eqnarray}
	\frac{\epsilon}{a} Q_{n,\gamma}^\ast(\boldsymbol{\theta}) + \left(1 - \frac{\epsilon}{a}\right) Q_{n,\gamma}^\ast(\boldsymbol{\eta}_n)
	\geq Q_{n,\gamma}^\ast(\boldsymbol{\theta}^\ast) 
	&\geq& \frac{1}{2}\epsilon^2 - \frac{1}{2}\parallel\boldsymbol{\eta}_n\parallel_2^2 - \Delta_n 
	\nonumber\\
	&\geq & \frac{1}{2}\epsilon^2 + Q_{n,\gamma}^\ast(\boldsymbol{\eta}_n) - 2\Delta_n.
	\nonumber
\end{eqnarray}
Taking infimum outside the ball $B_1(n)$, and using  (\ref{EQ:Ct.4}) and $\epsilon<a$,  we get for large enough $n$, 
$$
\inf\limits_{\parallel\boldsymbol{\theta} - \boldsymbol{\eta}_n\parallel_2>\epsilon} Q_{n,\gamma}^\ast(\boldsymbol{\theta})
\geq Q_{n,\gamma}^\ast(\boldsymbol{\eta}_n) + \frac{a}{\epsilon}\left[\frac{\epsilon^2}{2} - o_P(1)\right] 
> Q_{n,\gamma}^\ast(\boldsymbol{\eta}_n).
$$
Thus $\widehat{\boldsymbol{\theta}}_n$ lies within the ball $B_1(n)$ with probability tending to one, and using the asymptotic convergence in (\ref{EQ:Ct.3}) and Slutsky's theorem, we finally get
\begin{eqnarray}
	\boldsymbol{u}^T[\boldsymbol{Z}_n^T\boldsymbol{\Omega}_n\boldsymbol{Z}_n]^{-1/2} \left[ \widehat{\boldsymbol{\theta}}_n 
	+ n\lambda_n\boldsymbol{V}_n\widetilde{\boldsymbol{w}_0} \right]
	\mathop{\rightarrow}^{\mathcal{L}} N(0,1),
	\nonumber
\end{eqnarray}
and substituting $\widehat{\theta}_n = \boldsymbol{V}_n^{-1}\left(\widehat{\beta}_1 - \beta_{01}\right)$ we obtain the stated convergence.






\end{document}